\documentclass[namedreferences]{solarphysics} 
\usepackage[optionalrh]{spr-sola-addons} 
\usepackage{natbib}
\usepackage{graphicx}                    
\usepackage{color}                       
\usepackage{url}                         

\newcommand{\kms}{    {$\mathrm{km\,{s}^{-1}}$}}
\newcommand{\acc} {    {$\mathrm{m\,{s}^{-2}}$}}
\newcommand{\erg} {     {$\mathrm{erg\,{cm}^{-2}\,{s}^{-1}}$}}
\newcommand{\rs}  {     {$\mathrm{R_{\odot}}$}                }
\newcommand{\w}   {{\textit{W}}}
\newcommand{\pw}   {{\textit{P-W}}}

\begin{document}

\begin{article}

\begin{opening}

\title{On the Nature and Genesis of EUV Waves: 
A Synthesis of Observations from SOHO, STEREO, SDO, and \textit{Hinode}}

%
\author{Spiros ~\surname{Patsourakos} $^{1}$\sep
        Angelos ~\surname{Vourlidas}$^{2}$\sep
       }

%

%
  \institute{$^{1}$ University of Ioannina,
Dept Physics-Section Astrogeophysics,
GR 451 10 Ioannina,
Greece
                     email: \url{spatsour@cc.uoi.gr} \\ 
  $^{2}$ Space Science Division, Naval Research Laboratory, Washington, DC 20375, USA 
\\
             }

\begin{abstract}
  A major, albeit serendipitous, discovery of the \textit{SOlar and
  Heliospheric Observatory\/} mission was the observation by the
  \textit{Extreme Ultraviolet Telescope} (EIT) of large-scale Extreme
  Ultraviolet (EUV) intensity fronts propagating over a significant
  fraction of the Sun's surface. These so-called EIT or EUV waves are
  associated with eruptive phenomena and have been studied
  intensely. However, their wave nature has been challenged by
  non-wave (or pseudo-wave) interpretations and the subject remains
  under debate. A string of recent solar missions has provided a
  wealth of detailed EUV observations of these waves bringing us
  closer to resolving their nature. With this review, we gather the
  current state-of-art knowledge in the field and synthesize it into a
  picture of an EUV wave driven by the lateral expansion of the
  CME. This picture can account for both wave and pseudo-wave
  interpretations of the observations, thus resolving the controversy
  over the nature of EUV waves to a large degree but not
  completely. We close with a discussion of several remaining open
  questions in the field of EUV waves research.
\end{abstract}

%

\end{opening}

\section{Introduction}

One of the most important, as well as intriguing, discoveries of EIT 
\citep{eit} on-board the \textit{Solar and
Heliospheric Observatory} (SOHO) were the EIT or EUV waves
\citep{moses1997,thomp1998,thomp1999}.  These are
brightness fronts which propagate over significant
fractions of the solar disk, mostly over quiet Sun (QS) areas, at
speeds which can reach several hundred $\mathrm{km\,{s}^{-1}}$
(e.g. \opencite{thomp2009}).  EUV waves are associated with large-scale
eruptive phenomena like flares and coronal mass ejections
(CMEs). The sources of EUV waves lie within 
active regions (ARs).

A large statistical survey of EUV wave observations
from EIT by \citet{biese2002} showed a high degree
of correlation between EUV waves and CME onsets for
well-defined EUV waves; this was not the case
for flares.  Furthermore, \citet{chen2006}
studied a set of energetic flares ($\ge$ M GOES class)
and found that only the eruptive ones were associated with
EUV waves.  On the other hand, weak flares (A and B GOES class),
are often associated with EUV  waves, whenever they are 
eruptive. From all the above we conclude that the existence
of a CME is a strong condition for the occurrence of an EUV wave.

\begin{figure}
\centerline{\includegraphics[width=0.85\textwidth,clip=]{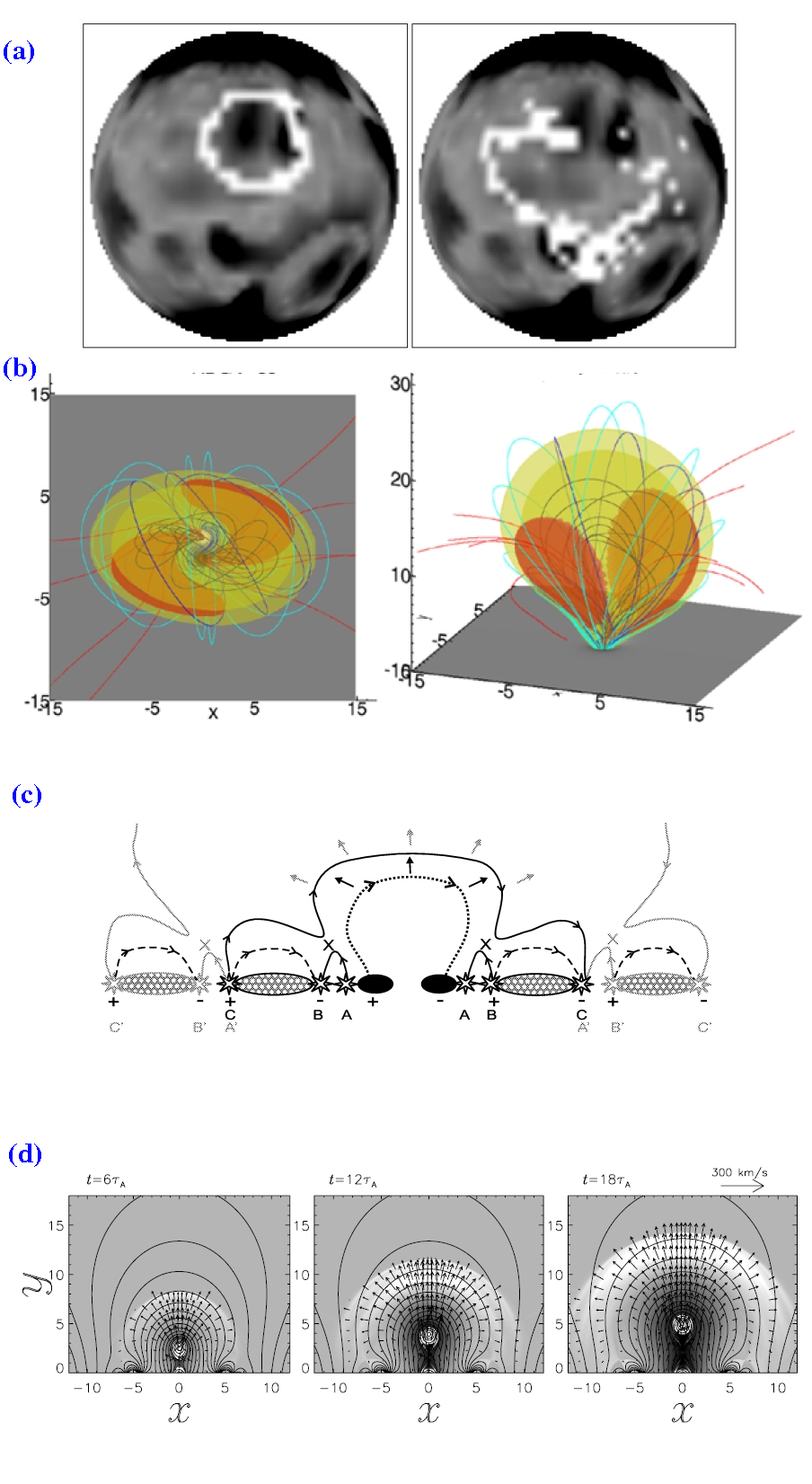}}
\caption{Proposed Physical Mechanisms for EUV waves.  MHD fast-mode
  wave, (panel (a), from \citet{wang2000}). Pseudo-wave - current shell
  (panel (b), from \citet{dela2008}), pseudo-wave - reconnection front,
  (panel (c), from \citet{attrill2007a}) ; hybrid pseudo-wave + MHD
  wave/shock,(panel (d), from \citet{chen2002}). Panels (a), (c) and  (d)
reproduced by permission of the AAS.}
\label{fig:wavemech}
\end{figure}

Several mechanisms have been suggested to explain the nature of EUV waves: 
(1) {\it ``true'' waves} (e.g. fast-mode)
waves), (2) {\it pseudo-waves} (e.g., compression fronts, current
shells, and reconnection fronts around and/or at erupting flux ropes
and (3) {\it hybrid}, i.e.  a combination of both wave and pseudo-wave
components. Examples from these mechanisms are given in
Figure~\ref{fig:wavemech}.

The {\it wave} interpretation asserts that EUV waves are ``true'' wave
phenomena; namely, a fast-mode wave that is (most likely) triggered
by a CME
\citep[e.g.,][]{thomp1998,thomp1999,wang2000,wu2001,ofmanthompson2002}. Such
waves have two attractive properties: (i) they can propagate
perpendicularly to the magnetic field and thus travel across the solar
surface and (ii) they are compressive waves and hence can be detected
in EUV images. Their typical speeds (200-400\kms) are in the range of
the anticipated fast-mode speeds over QS. Another suggestion in the
frame of the wave scenario, is that EUV waves are solitary waves,
i.e. solitons \citep{soliton}.

The {\it pseudo-wave} interpretation suggests that EUV waves are not
true wave phenomena, but rather the disk projection of the CME's
expanding envelope (\opencite{delaaul1999}).  In this envelope, the
plasma is been compressed by the flanks of the expanding CME flux rope
and/or heated and compressed in a current shell around or at the
surface of the erupting CME flux rope (to ensure current neutrality in
the former case) \citep[e.g.,][]{dela2000,dela2008}.  Another variant
of a pseudo-wave is that the CME flux rope laterally expands across
the solar surface with a series of magnetic reconnections between the
rope magnetic fields and QS magnetic fields of favorable
orientation. These, presumably low-energy, reconnections cause
transient brightenings which can give the appearance of an EUV wave,
when collectively perceived \citep{attrill2007a, attrill2007b}.

A third interpretation, the {\it hybrid} wave,  tries to bridge the previous two
opposing views, by first acknowledging that multiple bright fronts can
be sometimes seen during events with EUV waves. Although this fact is
quite obvious to anyone familiar with the observations, it is
not always discussed clearly in the literature. According to the
\textit{hybrid\/} interpretation, there are two 'EUV waves'. A sharp
outer front consistent with a fast-MHD shock and considered the
coronal counterpart of the Moreton wave, and an inner diffuse front
consistent with a pseudo-wave and considered the 'typical' EUV wave
\citep{chen2002,chen2005,zhukov2004,cohen2009,pomoell2008}. The latter
is the key weakness of this interpretation, as we will
discuss later. While accepting the existence of two
fronts on theoretical grounds, the hybrid interpretations tend to dismiss the
observational detection of two fronts for a given event in favor of a
single front associated with the expanding CME (or the
surrounding field). As we will discuss in Section 12, this is not the
complete picture. The two fronts exist, they are observed, and recent
theoretical work sheds a much clearer light in this \textit{hybrid\/} view
\citep{downs2011}.

At this point we should emphasize that observations reveal a large
diversity of moving features in association with the EUV wave. This
becomes evident when the image cadence, observational lines of sight,
and temperature coverage are increased throughout the solar
atmosphere. The long list of observed features includes expanding
loops, mass flows, core and extended dimmings, stationary
brightenings, deflections, and oscillations of ambient coronal
structures. These manifestations occur in tandem to the large-scale
front which is essentially the EUV wave.  Moreover, EUV waves are
  frequently associated with phenomena observed in other wavelength
  domains such as Moreton waves in the chromospheric H$\alpha$
  \citep[e.g.,][]{thomp2000,warm2001} and He I 10830 \AA \, lines
  \citep[e.g.,][]{vrsnak2002}, and in the corona in Soft X-rays (SXRs)
  \citep[e.g.,][]{khan2002,hudson2003,warm2005} microwaves
  \citep[e.g.,][]{warm2004,white2005} and in the metric-range
  \citep[e.g.,][]{vrsnak2005}. 

We also note that there is a hierarchy of EUV waves.  The higher
cadence and sensitivity of the EUVI and AIA observations showed the
existence of small-scale waves which are associated with and are
probably triggered by small-scale eruptions, like small erupting
filaments, \citep[e.g.,][]{innesminiwave2009, podla2010,
  zhengminiwave2011, cyclones}.  The speeds of these "mini-waves"
range between 10 and 250\kms and the waves leave small dimmings
behind them. However, they do not reach the global scales that
"ordinary" EUV waves since they travel over distances of the order of
around 100 Mm only. In this paper, we will deal exclusively with {\it global\/} EUV waves,
i.e. well-developed, clearly visible (at least initially) propagating
fronts which reach distances of a significant fraction of the solar
radius.  This is in our opinion an objective criterion because it does
not depend on the derived speed of the propagating disturbance which
seems to be sometimes a function of image cadence. For example, EIT captured
only the slower of EUV waves, due to its low-cadence, as was revealed
by the higher cadence STEREO/SECCHI \citep{secchi} 
observations \citep{long2008,veronig2008wave}.

EUV waves comprise a very active field of coronal research which is
characterized by occasional controversy and intense
debate. Consequently, the subject has been reviewed extensively over
the years. Recent reviews on EUV waves can be found in
\citet{warm2007}; \citet{willsrev2009}; \citet{warm2010};
\citet{galla2011} and \citet{zhukov2011}. We first focus on multi-viewpoint observational
  results (this is a 'Sun-360' Topical Issue, after all), discuss
  energetics, and finally propose a top-level synthesis of the CME-EUV
  wave interplay that, we believe, accounts for the majority of the
  observations and can resolve past controversies. Of course, we
  update the field with as many recent publications as we could. We
  start with a review of the kinematics and coronal interactions of
  the waves, proceed to the thermal and 3D structure, and then discuss
  associated phenomena, such as brightenings, ripples, etc, that can
  lead to confusion. We then turn our attention to two
  rarely-discussed subjects; namely, the spectroscopic observations
  and energetics of EUV waves. In Section 10-11, we synthesize the
  current knowledge on this phenomenon by first discussing the genesis
  of EUV waves and then describe a picture of the CME-EUV wave
  connection that seems to be consistent with the majority of the
  observations. We conclude in Section~12 with a list of open
  questions and possible areas of research in near-future. We
  concentrate on recent observations acquired by space missions, such
  as SOHO, Hinode, STEREO and SDO. We use the term EUV wave throughout
  the paper regardless of the physical interpretation ("true" wave or
  pseudo-wave) of the phenomenon.

\section{Kinematics, Amplitudes and Dispersion}
\label{sec:kin}
Arguably the best studied property of EUV waves is their
kinematics. The steady increase in image cadence, from 720 s with SOHO/EIT
to 150 s with EUVI on STEREO/SECCHI, and finally to 12 s with SDO/AIA,
is the key factor for improving our understanding.  'Point-and-click',
semi-, and fully-automated methods are used to determine the
time-distance curve ($t-d$) or ground-track of the wave in one or more
angular sectors. Then, simple numerical derivation or fittings of the
($t-d$) curves with various functions provide the speed and
acceleration profile of the wave.

\begin{figure}    
                                
   \centerline{\hspace*{0.015\textwidth}
               \includegraphics[width=0.49\textwidth,clip=]{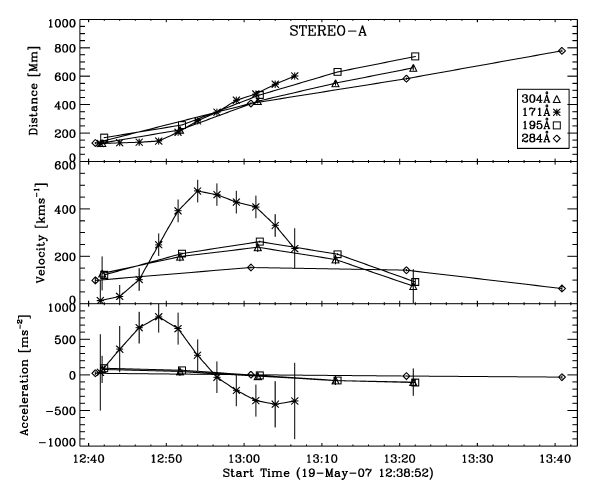}
               \hspace*{-0.01\textwidth}
               \includegraphics[width=0.49\textwidth,clip=]{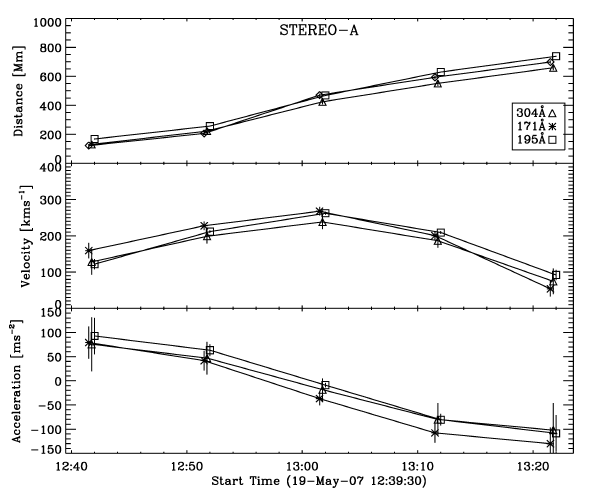}
              }
\caption{Time evolution of distance, speed
and acceleration for an EUV wave which took place
on 19 May 2007. Left panel has 171 channel observations
at 2.5 minute cadence while right panel at a reduced 10 minute
cadence. From \cite{long2008}. Reproduced by permission of the AAS.}
\label{fig:long2008}
\end{figure}

The first STEREO/EUVI detailed kinematic observations of an EUV wave
on 19 May 2007 by \cite{long2008} and
\cite{veronig2008wave} showed that the lower cadence of EIT {\it
  underestimated} the initial speeds of the waves. The higher EUVI
cadence revealed significant {\it deceleration} of the first
wave fronts from $\approx$ 400\kms to 200\kms in a matter of almost
10 minutes; after this interval the wave was traveling at an almost
constant speed of about 200\kms (left panel of Figure
\ref{fig:long2008}).  Lower cadence observations would have missed
significant part of the deceleration phase (right panel of Figure
\ref{fig:long2008}). For higher initial wave speeds ($>$400-500\kms) even the EUVI cadence is inadequate. The first EUV wave
observations by AIA \citep{aia}
 showed examples of EUV waves with very high
initial speeds (650-2000\kms) undergoing large decelerations, up to -2.0$\times {10}^{3}$ \acc \, 
\citep{chenaia2011,ma2011,kozarev2011,liu2000,cheng2012,oo2012}.

  Compilation of the kinematics of other events observed by EUVI and
  AIA showed either waves experiencing significant deceleration in
  their early stages \citep[e.g.,][]{kozarev2011,long2011a,ma2011,muhr2011, warm2011} or 
  waves with $\approx$ constant speeds
  \citep[e.g.,][]{kienquad2009, ma2009, wavestereo, quadwave, temmer2011, kienhomol2011, liu2010, long2011a,
warm2011}.
  Irrespective of their initial speeds or deceleration profiles, these
  waves ended up travelling within a rather narrow speed range of
  180-380\kms which is consistent with the fast mode speed over the
  quiet Sun \citep[e.g.,][]{wang2000, wu2001, warm2005alf, cohen2009,
    schmidt2010, downs2011, zhao2011}.  Therefore, the kinematics of
  the observed waves are consistent with fast-mode waves. Such
  waves are initially driven and even shocked sometimes (for
  the fast and decelerating events) with their initial speeds
  reflecting the speed of the driver and not the characteristic speed
  of the medium where they propagate.  Note here that all these events
  correspond to "truly" global waves since they cover distances
  350-850 Mm, or conversely 0.5-1.3 $\mathrm{R_{\odot}}$; these waves
  were also "well-observed" events showing clear evidence of
  propagation of well-defined fronts.

   At this point it will be useful to discuss some of the key
    properties of both linear and non-linear fast mode waves, and
    their kinematic behavior in particular.  Detailed discussions on
    this topic can be found in
    \cite{mann1995,vrsnak2000,warm2007}.  Waves with large
    amplitudes, shocks being a special case, cannot be treated
    linearly and their speeds are always higher than the ambient
    fast-mode speed profile.  Their kinematics would depend on the
    wave amplitude. When energy input ceases (i.e. blast-wave), the
    wave decelerates when traveling in a constant fast-mode speed
    medium since its amplitude is decreasing due to profile broadening
    and geometrical expansion. On the other hand, when a non-linear
    wave is still driven, due to the expanding CME acting as a piston,
    for example, its amplitude increases, and therefore it
    accelerates. When the amplitude of the disturbance is
    small, we have linear waves travelling at the characteristic
  speed of the medium.  Therefore, linear waves would follow the
    fast-mode speed profile of the ambient medium and would travel at
    constant speeds for uniform fast-mode profile, as expected for
    propagation over QS areas. However, linear waves could also
    experience acceleration or deceleration when they cross the
    boundaries between regions with strong fast-mode speed gradients,
    like QS, coronal holes or ARs (see Section 3). In conclusion, both
    linear and non-linear fast-mode waves could exhibit constant
    speed, accelerating and decelerating kinematic profiles and should
    reflect or dissipate at coronal hole or AR boundaries.

\begin{figure}
\centerline{\includegraphics[width=0.65\textwidth,clip=]{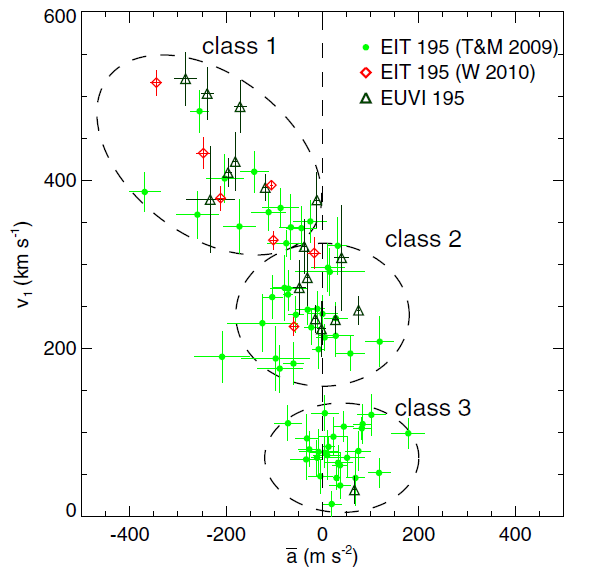}}
\caption{Statistical relationship between initial EUV wave speed
and average acceleration for 17 waves observed by EUVI and
61 waves observed by EIT. From \cite{warm2011}. Credit: Warmuth, A.,
Mann, G., A$\&$A, 532, 151, 2011, 
reproduced with permission \copyright ESO.}
\label{fig:warm2011}
\end{figure}

\begin{figure}
\centerline{\includegraphics[scale=0.6,clip=]{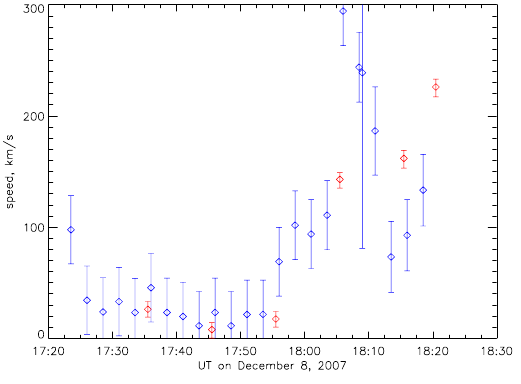}}
\caption{Time-speed plot for an EUV wave which took place on 8 December 2008
exhibiting erratic kinematic behavior.
From \cite{zhukov2009}.}
\label{fig:zhukov2009}
\end{figure}

This behavior has been put into a broad context
with the recent extensive statistical study of \cite{warm2011}
where the kinematics of a comprehensive set of
EUV waves (61 observed by EIT and 17 observed by EUVI)
has been used. Most of the EIT observations 
were from a catalog of 176 EUV waves compiled by \cite{thomp2009}.
Figure \ref{fig:warm2011} shows  
the initial wave speed 
against the  mean  wave acceleration. 
For initial wave speeds exceeding roughly 320\kms  there is a clear
trend that faster waves experience stronger decelerations (class 1 in Figure \ref{fig:warm2011}).
Waves with intermediate speeds ($\approx$ 170--320\kms) are characterized by small magnitude accelerations
or decelerations which is consistent 
with $\approx$ constant speed (class 2 in  Figure \ref{fig:warm2011}).
Finally, very slow waves (speeds $<$ 120\kms)
exhibit small accelerations/decelerations (class 3 in  Figure \ref{fig:warm2011}).
It is unlikely that the fast-mode wave
interpretation applies in these cases since their speeds are smaller or
of the same order as the coronal sound speed, 
and the fast-mode speed depends on the quadratic sum of the
sound and the Alfv\'{e}n speeds. These slow waves are consistent
with pseudo-wave interpretations although a
slow-mode wave travelling at almost $ 90 ^{\circ}$ 
with respect to the ambient magnetic field is another possibility \cite{podla2010}. 
Another clue to the non-wave nature of the slow EUV waves is 
that they do not cover big distances during their lifetimes, so they cannot qualify as global waves. 
One example of a slow pseudo-wave was reported by \cite{zhukov2009}.
The measured time-speed profile (see Figure \ref{fig:zhukov2009}) 
was not smooth but exhibited a series accelerations and decelerations.
In reality, the observed "EUV wave" was just the footprint of an erupting
filament undergoing rotation. This example highlights the importance
of analyzing the wave kinematics in order to deduce the nature
of the observed waves.

The study of the {\it perturbation profiles (pp)} of EUV waves is
another powerful tool.  Although different definitions can be
found in the literature, a {\it pp} profile essentially measures the
intensity distribution across the wave, $I(r,t)$, with respect to the
distribution, $I(r,t_{0})$, at a pre-wave moment
$t_{0}$. It is  $ pp
\propto I(r,t)/I(r,t_{0})$ , where $r$ is the distance from a fixed
initiation point of the wave.  For any given time, the maximum and the
$FWHM$ of the {\it pp} are deduced by fitting the {\it pp} 
with a Gaussian profile, for example.  

The {\it pp} analysis of EUV waves resulted in several interesting
findings \citep{veronig2010dome,long2011a,muhr2011}. The wave $FWHM$ is
generally an {\it increasing} function of distance.  At the same time,
the wave amplitude $A$ (i.e. peak value of the {\it pp}) 
  {\it decreases} with time.  The {\it pp} analysis of EUV waves and
  of the associated Moreton waves by \citet{warm2010} showed they both
  exhibit profile broadening and amplitude decrease. For
some events, the integral of {\it pp} over distance which is
proportional to $A \times FWHM$ is constant or decreases with
distance.  All the above suggest that the observed disturbances are
consistent with freely-propagating (blast) waves since the total
energy is either constant (constant $A\times FHWM$) or decreasing
(decreasing $A\times FHWM$). The pulse broadening points also to a
freely-propagating wave.  Indeed, \citet{grech2011} found that a 3D
model of a blast wave propagating in a medium with density
stratification was broadly consistent with the ground track of
17 January 2010 (see also \citet{veronig2010dome}). Finally, MHD
modeling of rotating sunspots by \citet{selwa2012} showed
that dome-like structures, similar to what is observed, could
be generated.

A recent study of a wave which took place on 14 August 2010 and was
observed by AIA allowed to deduce the wave dispersion characteristics
with ultra-high cadence and at multiple channels
(\opencite{long2011b}).  The AIA observations showed that the wave width
increases with time in all channels while its magnitude was
decreasing. By treating the {\it pp} of the waves as a linear
combination of sinusoidal waves within a Gaussian envelope,
\citet{long2011b} found a dispersion rate of the pulse of 8-13
$\mathrm{{Mm}^{2}{s}^{-1}}$. The dispersive nature of the observed
wave is a strong indication for its wave nature.  Wave dispersion is
at odds with solitary waves.

\begin{figure} 
\centerline{\includegraphics[width=0.65\textwidth,clip=]{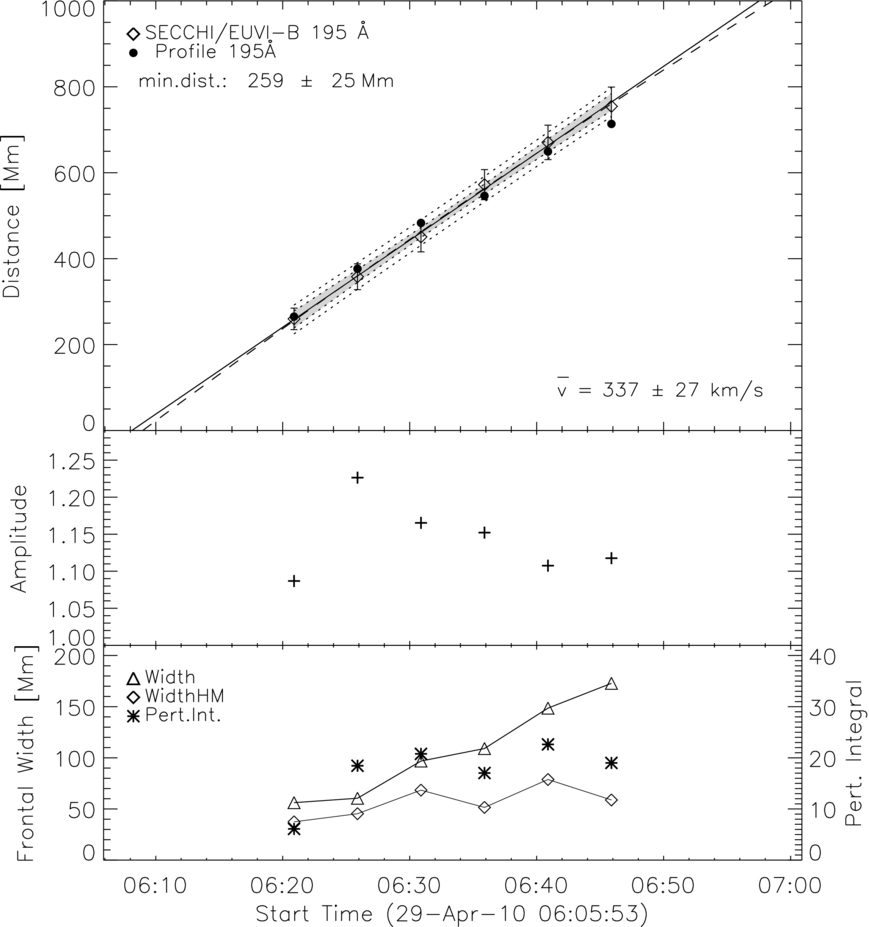}}
\caption{Analysis of perturbation profiles of an EUV wave which took place
on 10 April 2010.
Time evolution of wave distance (upper panel), wave amplitude (middle panel)
and wave width and perturbation integral (lower panel). From \cite{muhr2011}.
Reproduced by permission of the AAS.}
\label{fig:muhr}
\end{figure}

\citet{kienhomol2011} presented the first observations of {\it
  homologous} EUV waves; within a period of 10 hours during April
28-29 2010 four EUV waves were launched from the same AR and along
the same direction.  This basically ensures that these waves
propagated over more or less the same background coronal conditions
(i.e., plasma $\beta \approx$ constant). It was found that the faster the
wave the larger the corresponding maximum compression ratio $X_{c}$
($\propto {(I/I_{0}})^{1/2}$).  \citet{kienhomol2011} calculated the
corresponding magnetosonic Mach numbers $M_{ms}=f(X_{c},\beta)$ from
the observed compression ratios for each event and found they were
strongly correlated with the (linear) wave speed. This result provided strong
support that the observed waves were indeed fast-mode
shocked (non-linear) waves.

\section{Interactions with the ambient corona: 
reflections, transmissions and oscillations}
\label{sec:inter}
MHD waves, as any type of wave, must follow the basic rules of optics.
This means that at least part of the wave could be reflected away from
places of strong gradients in the characteristic speed of
the medium where they propagate. For the case of the fast-mode speed,
strong gradients are expected at the interfaces between QS and coronal
holes and ARs where the fast-mode increases from few hundred \kms\ to
several hundred or even thousand \kms\ \citep[e.g.,][]{wang2000, schmidt2010}.
Wave transmission through a coronal hole could
occur when a resonance between the coronal hole and the incoming wave
takes place \citep{schmidt2010}. The EUV wave simulations of
\citet{wang2000} and \citet{schmidt2010} showed wave reflection from
the boundaries of coronal holes.

\begin{figure}
\centerline{\includegraphics[scale=0.6,clip=]{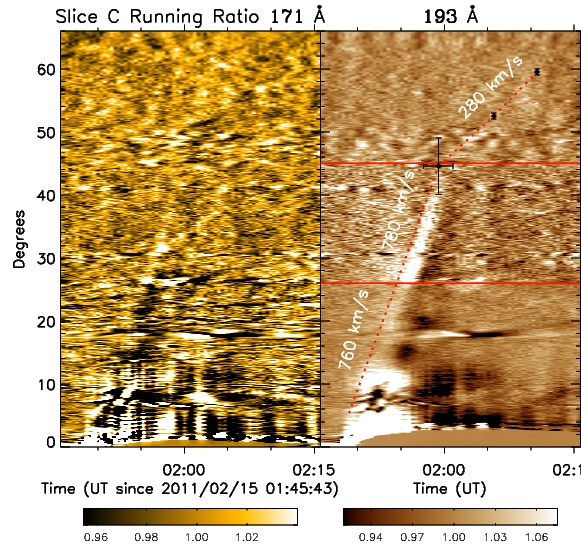}}
\caption{Combined AIA-EUVI observations of wave reflection and transmission.
Time-angle plot along a given direction of
an EUV wave which took place on 15 February 2011. Red dots are
the wave ground tracks and the two horizontal red lines
define a coronal hole. From \citet{oo2012}. Reproduced by permission of the AAS.}
\label{fig:oo}
\end{figure}

The EUVI observations of the 19 May 2007 wave by
\citet{gopalrefle2009} showed evidence of wave reflection from a small
equatorial coronal hole. The speed of the reflected wave was similar
to the speed of the incoming wave.  The reported reflection was put into some question by
\citet{attrill2010}, based on the running-difference images used in
the \citet{gopalrefle2009} analysis.  It is true than
running-differences could lead to some confusion, particularly when
one is looking at reflections since these occur over regions which
are already perturbed by the incoming wave. However evidence of wave
reflection for this event is also seen in the direct images.

\begin{figure}
\centerline{\includegraphics[width=0.85\textwidth,clip=]{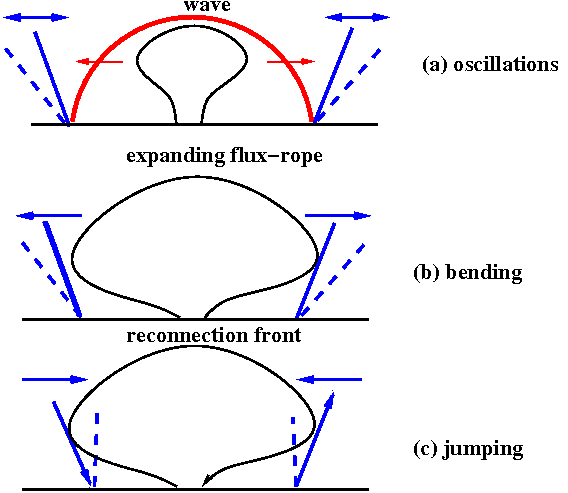}}
\caption{Expected interactions between EUV waves and ambient coronal
  structures in the frame of the various physical mechanisms proposed
  for EUV waves. The mushroom-shaped object is the laterally and
  radially expanding CME structure. The inclined solid blue lines
  represent ambient coronal structures before the interaction with the
  wave or CME. The inclined dashed blue lines represent the new
  locations of the ambient coronal structures after the
  interaction. The blue arrows show the direction of motion of the
  ambient structures induced by the interaction.  Upper panel: Wave
  mechanism.  The expanding CME launches a wave (red) which sets the
  ambient coronal structures into oscillations. Middle panel:
  CME Current-shell, CME-compression
  front mechanism. There is no wave. The expanding CME pushes
  the ambient coronal structures \textsl{continuously\/} as it
  expands. No oscillations should be produced. Lower panel:
  Reconnection mechanism. No wave is produced. The expanding CME
  reconnects with the ambient structures of opposite polarity which
  should then \textsl{``jump''} inwards. No oscillations should be
  observed.}
\label{fig:defle}
\end{figure}
The angular separations between the two STEREO and the SDO spacecraft
provided for the first time a 360-degree coverage of an EUV wave which took place
on 15 February 2011 \citep{oo2012}. The source AR was close to the central meridian and wave tracks travelling both towards
the southeast and the southwest were reflected off an extended south pole
coronal hole. In addition, part of the wave was {\it transmitted} through
the coronal hole rather than reflected (see Figure \ref{fig:oo}).  The
wave approached the hole with a speed of 760\kms; the transmitted
part traveled within the coronal hole at a speed of 780\kms, once the
wave either reflected off or exited through the coronal hole it traveled at a slower
speed of around 280\kms.  This kinematic behavior fits 
nicely with a wave interpretation. An initially driven wave (incoming
wave) reaches the coronal hole and part of it is reflected by the
strong fast-mode speed gradient of the coronal hole and part is
transmitted through it. The faster propagation speed within the
coronal hole correlates with the higher fast-mode speeds within
coronal holes; the slower speeds of the reflected and transmitted
waves which travel over QS are consistent with typical QS fast-mode speeds.
The observed very fast wave transmission through a coronal hole makes
it possible that pre-SDO observations could have missed similar
effects in other events (i.e.  the wave transit time through the hole
is only 5 minutes). Wave reflection and transmission through a coronal
hole are hard to reconcile with pseudo-waves since CMEs never propagate into
coronal holes for example. Transmission into coronal holes is
sometimes observed for Moreton waves as well \citep{veronig2006}.

\begin{figure}
\centerline{\includegraphics[scale=0.6,clip=]{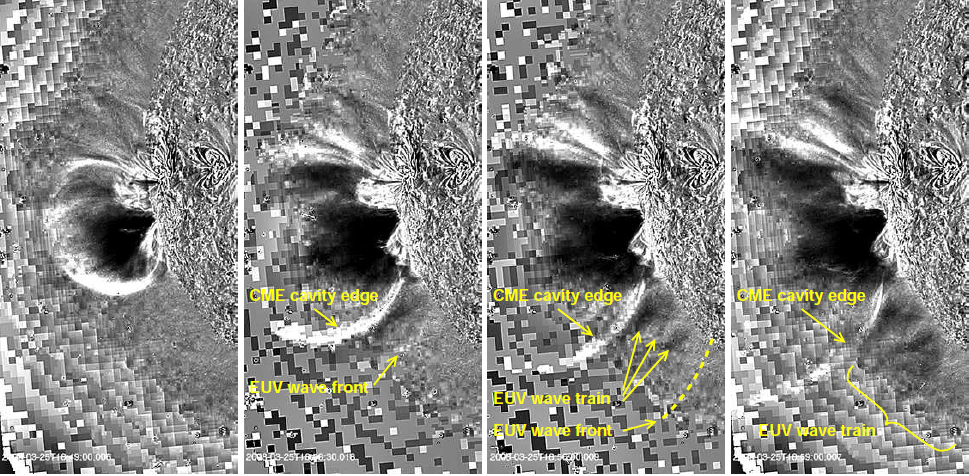}}
\caption{Deflections of ambient coronal structures from the impact
of an EUV wave which took place on 25 March 2008. From \citet{bubblestereo}.
Credit: Patsourakos, S., Vourlidas, A., Kliem, B., 522, 100,
2010,
reproduced with permission \copyright ESO.}
\label{fig:wave_defle}
\end{figure}

Motions in the form of deflections/oscillations of ambient coronal
structures are another important indicator of the nature of EUV waves.
Looking at any high cadence off-limb movie of EUV waves 
\citep[e.g.,][]{quadwave,bubblestereo}
one sees evidence of
oscillations of ambient coronal structures which are set up once the
wave impinges on them (e.g. Figure \ref{fig:wave_defle}). The outermost
oscillating structures roughly outline the locations of the wave
front. To illustrate this effect we included as electronic
supplement a 193 AIA channel
base-ratio movie of the event of the 13 June 2010.  In the southern
part of the eruption we note an area growing with time which exhibits
oscillations of off-limb structures. These are manifested as
alternating black-and-white stripes at any given location.  The most
natural explanation for these oscillations is that a true MHD wave
impinges on the ambient coronal structures and sets up an oscillation
(e.g. panel (a) of Figure \ref{fig:defle}).  Similar deflection
phenomena have been observed further away as streamer deflections with
coronagraphs in  connection with CME-driven shocks 
\citep[e.g.,][]{gosling1974, sheeley2000defle,vourlidasshock}.
\citet{vrsnak2006} and \citet{tripathi2007} reported
streamer deflections in direct temporal and spatial association
with EUV waves.

On the other hand, if an expanding flux rope were responsible for
these deflections, it should bend the ambient coronal structures
rather continuously along the direction of its lateral expansion as it
rolls over them (panel (b) of Figure \ref{fig:defle}). We would not
observe oscillations in that case. Finally, a reconnection front
(panel (c) of Figure \ref{fig:defle}) would cause opposite directed
``jumps'' (which could be interpreted as oscillations). Once again,
the observations of oscillations of ambient structure support a
wave-interpretation for EUV waves.

Other phenomena trigerred by EUV waves are kink-like oscillations of
disk filaments \citep{her2011} and coronal loops
\citep[e.g.,][]{willsthomp1999, ma2011kink}.  The velocity of these
transverse oscillations is in the range of 5-50\kms, the same range
as filament oscillations trigerred by Moreton waves
\citep[e.g.,][]{gil2008}; for a review
on filamenent/prominence oscillations and their
dissipation mechanisms the interested reader can refer
to \citet{promosc}.
More recently, AIA observations of the 2011
June 7 event by \citet{li2012} showed evidence of {\it secondary}
waves triggered in nearby ARs or individual loop-like structures when
the main wave hits them. Using the observed velocities, and typical
masses and densities for filaments and coronal loops, one can estimate
the kinetic energy associated with the oscillations
\citep[e.g.,][]{ballai2005}.  The deduced energies are rather small,
in the range of nanoflares, and set a lower limit on the total energy
of EUV waves.

\section{Thermal Structure}
\label{sec:thermal}
The first truly multi-thermal observations of EUV waves were achieved
by EUVI, when the same EUV wave was observed in 4 different channels
(171, 195, 284, 304) with cadence from 2.5-20 minutes
\citep[e.g.,][]{long2008, veronig2008wave, wavestereo}.  The main
result was that EUV waves are best observed in the 195 channel, which
has a peak response temperature of about 1.5 MK. They are weaker in
other channels.

\begin{figure}
\centerline{\includegraphics[width=0.9\textwidth,clip=]{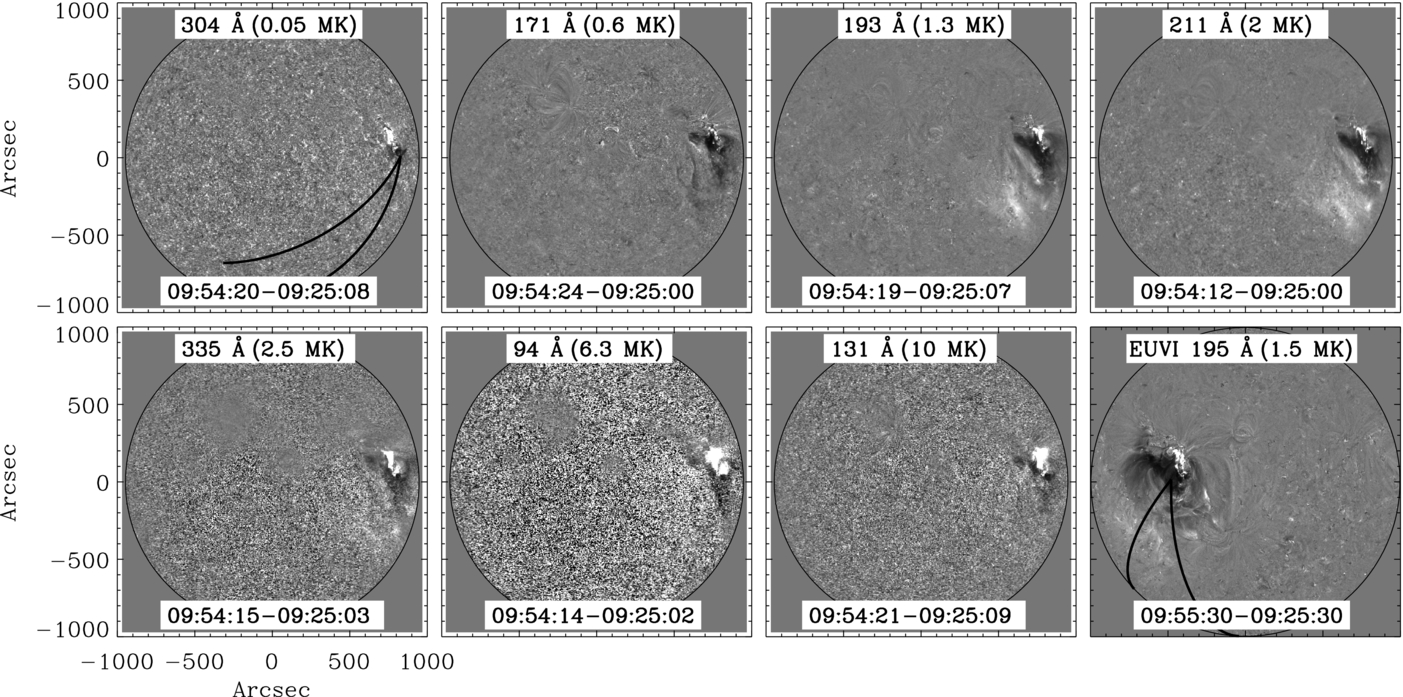}}
\caption{AIA observations of an EUV wave 
which took place on 14 August 2010
in several EUV channels.  From \citet{long2011b}. Reproduced by permission of the AAS.}
\label{fig:multit}
\end{figure}

AIA improved the study of the thermal structure of EUV waves
by supplying observations at 12 s cadence in seven channels (94,
131, 171, 193, 211, 335, 304) sampling the transition region (0.08
MK), warm (1-3 MK) and flaring corona (6-10 MK). The AIA observations
of \citet{liu2010}, \citet{kozarev2011}, \citet{long2011b},
\citet{ma2011} and \citet{schriwave2011} showed that EUV waves are
best seen in the 193, 211, 335 channels which implies temperatures in
the range 1.0-2.5 MK (see for example Figure \ref{fig:multit}).  On
the other hand, EUV waves in the 171 channel (the cooler coronal
channel) are sometimes manifested as intensity {\it depletions}, i.e.
darkening, in contrast to the other coronal channels where wave fronts
are bright.  The 171 observations of intensity depletions
associated with EUV waves imply
that observations of a faint bright front
in a 304 channel imply that the 304 signal comes from the corona
(i.e., the Si XI line at 303.32 \AA \, in the 304 channel) and not from the transition region
(i.e., He II 304 \AA \, line ) as discussed in \citet{wavestereo} and \citet{long2011b}.

\citet{schriwave2011} modeled the intensity changes
associated with an EUV wave on 15 February 2011  using
the temperature response functions 
of AIA and the assumption 
that the observed variations were due only to
adiabatic compression. They found that
the data were consistent with mild plasma heating
and compression, a type of plasma ``warming''.
The estimated density and temperature increases
were $\approx$ 10 $\%$ 
and $\approx$ 7 $\%$ respectively, while
the "allowed" temperature range for the wave plasma was 1.2-1.8 MK.

\citet{kozarev2011} and \citet{ma2011}
analyzed the thermal structure of an EUV wave on 13 June 2010. 
This event was associated with a type II metric radio shock. \citet{kozarev2011}
used the AIA intensities recorded in several coronal channels
to perform Differential Emission Measure (DEM) analysis before and during
the EUV wave. They found that during the wave the DEM is increasing for
temperatures roughly exceeding the peak temperature of the pre-event DEM
($\approx$ 1.8 MK). This implies both plasma heating and compression.
Assuming no temperature change, they found a {\it lower}
limit for the density increase in the range 12-18 $\%$.
\citet{ma2011} deduced from the
dynamic radio spectrum of the associated shock a
compression ratio (1.56) and from the AIA data 
the wave speed. Feeding these parameters to
the jump conditions for a perpendicular MHD shock
they found down-stream plasma heating to $\approx$ 2.8 MK.
Using this temperature they found ionization time-scales
roughly consistent with observed timescales of the EUV
wave in the various AIA channels. 
This suggests that the observed EUV wave was consistent with a shocked wave.

\section{3D Structure and Relationship with CMEs}
\label{sec:3d}

Multi-viewpoint observations using the two STEREO spacecraft or
combinations of STEREO with SOHO or SDO can provide important
geometrical characteristics of EUV waves as well as the wave--CME
association in the 3D space including their lateral extensions.

Triangulations using STEREO data of an EUV wave which occurred on 7 December 2007
derived a wave front height of $\approx$ 90 Mm
(\opencite{wavestereo}).  This is comparable to the coronal
scale-height (70 Mm) at a temperature of 1.5 MK, which is the
characteristic temperature of the EIT and EUVI 195 \AA \ channels and
of the AIA 193 \AA\, channel. It may be the reason why EUV
  waves are usually best observed at this wavelength.  The above
height is indeed an ''emission'' height, i.e.  the height from which
the bulk of the wave emission originates. Wave emissions originating
from higher altitudes will be weaker and possibly invisible
(\opencite{robbre2009}) since the EUV emission has a strong dependence
on density which falls off very rapidly with height. These arguments
provide a strong constraint on the physical mechanism(s) of EUV waves
because they must act at the base of the corona irrespective of their
nature (wave or pseudo-wave) in order to give an observable signature.

\begin{figure}   
\centerline{\hspace*{0.015\textwidth}
\includegraphics[width=0.49\textwidth,clip=]{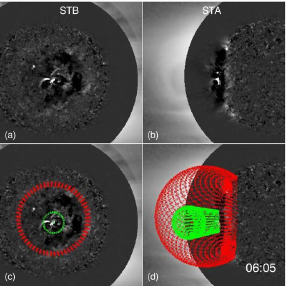}
\hspace*{-0.01\textwidth}
               \includegraphics[width=0.49\textwidth,clip=]{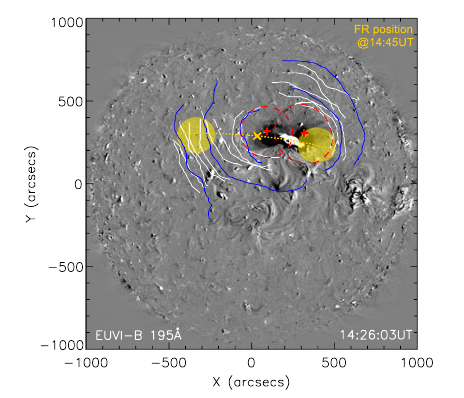}
              }
\caption{SECCHI multi-viewpoint observations and geometrical
modeling of  EUV waves and comparison with the associated CMEs. 
Left panel from \citet{quadwave} (13 February 2009 event)
 and right panel from \citet{temmer2011} (26 April 2008 event). Left panel
reproduced by permission of the AAS.}
\label{fig:3dwave}
\end{figure}

3D geometrical modeling of the EUV wave envelope and of the associated CME as seen
in the inner corona in the EUV or in white-light (WL) with
coronagraphs (e.g., COR1 on SECCHI) casts light into the wave-CME
relationship \citep{wavestereo,quadwave,temmer2011}.  For this task, widely separated views of the EUV
wave and of the CME are used. Ideally one view is off-limb or close to
the limb and the other is on-disk.  The 3D forward geometrical model
of \citet{therni2009} is used to obtain a 3D fit of the EUV wave and
of the associated CME which are then projected onto the solar disk
(e.g., Figure \ref{fig:3dwave}).  The results reveal a {\it spatial offset} and {\it size
  disparity} between the projections of the EUV waves and their associated CMEs which
indicates two possibly related (the wave running ahead of
the CME) but {\it different} entities.  Such results are clearly
inconsistent with pseudo-wave models for which wave and CME are the
same thing by definition.

\begin{figure}
\centerline{\includegraphics[width=0.99\textwidth,clip=]{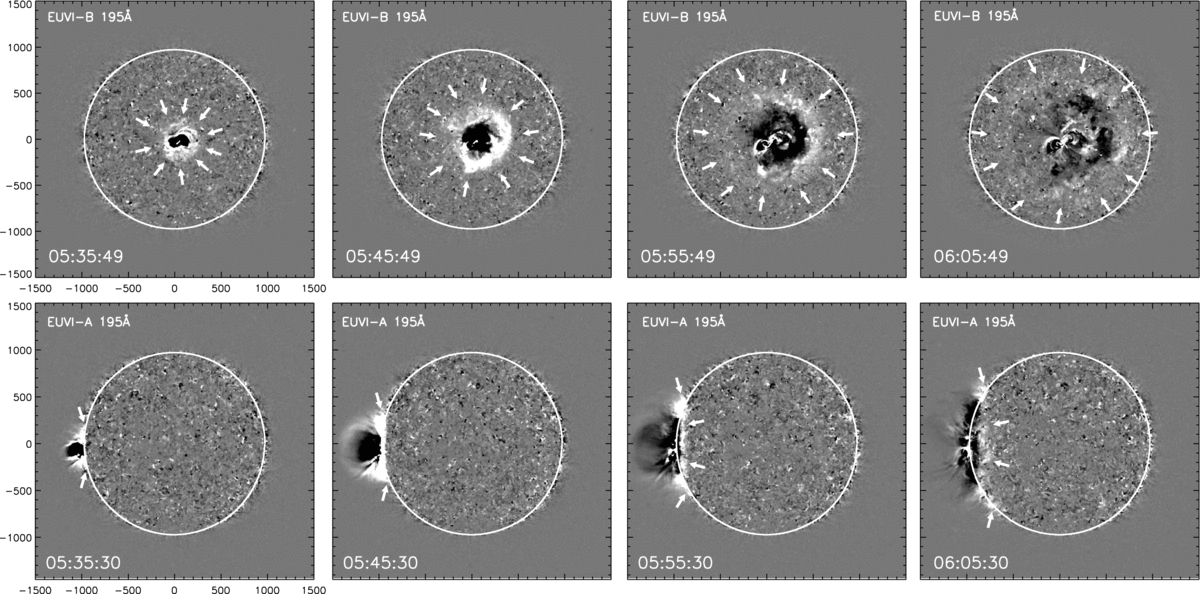}}
\caption{STEREO quadrature observations of an EUV which took place on
  13 February 2009. Upper panel STEREO A
  observations with an EUV wave and associated CME view from above and
  lower panel STEREO B observations with an EUV wave and associated
  CME view from the side. Median-filtered running difference images are shown.
From \citet{kienquad2009}.
Reproduced by permission of the AAS.}
\label{fig:over}
\end{figure}

\begin{figure}
\centerline{\includegraphics[width=0.6\textwidth,clip=]{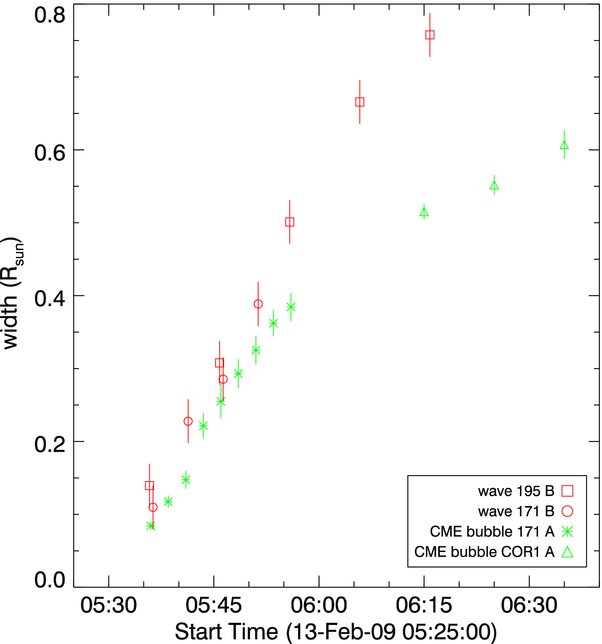}}
\caption{Distance time of the EUV wave (red symbols) from a disk-view and 
associated EUV-WL CME (green symbols) from an off-limb view 
for the quadrature observations of 13 February 2009. From 
\citet{quadwave}. Reproduced by permission of the AAS.}
\label{fig:quad}
\end{figure}

The ultimate test for the relationship between EUV waves and CMEs
became possible with the availability of {\it quadrature} STEREO
observations of an event which took place on 13 February 2009
\citep{kienquad2009,quadwave}, when the two STEREO
spacecraft had a $90\,^{\circ}$ separation.  The source AR was located
at disk center from STEREO B and at the east limb as seen from STEREO
A (Figure \ref{fig:over}). This is an ideal configuration for
measuring the wave (disk view) and CME kinematics
(limb-view) simultaneously (Figure \ref{fig:quad}).  It was found that 
the wave and CME were initially co-spatial but they decoupled quickly with
the wave detaching away from the CME flanks. The same conclusion was
reached by \citet{kienquad2009} from the analysis of off-limb maps of
the event taken at several heights. They found that the start of
lateral expansion of the EUV CME marked the initiation of the on-disk
wave and that the expansion started at a height of $\approx$ 90 Mm.  The
observed behavior suggested an initially driven disturbance
which eventually became a freely propagating MHD wave travelling at
around 250\kms \citep{kienquad2009, quadwave}. These results, 
especially the decoupling of the wave from the CME front, have been 
recently
verified with high cadence AIA observations in several events 
\citep{ma2011,cheng2012}.

Therefore, the CME front should follow the EUV wave front. But
  how can these results be reconciled with past reports of the exact
  opposite behavior \citep{vrsnak2006,warm2010} or with the
  claims of a single front \citep{attrill2009,dai2010}? In a
  nutshell and with the benefit of hindsight, these conflicting
  reports were cases of front misidentification due to poor wavelength
  and temporal coverage. There are only two reports of a CME front
  running ahead of an EUV wave (8 August 1998, 3 November 2003) and
  both of them were very poorly observed  lacking LASCO or EIT
  observations. Hence, in the case of the 3 November 2003  event, the CME
  front is assumed to be the soft X-ray front while the EUV wave is
  associated with the Moreton wave in H$\alpha$. But there is no
  independent proof that the CME and soft X-ray fronts are the
  same. Besides, there was a lot of SXR activity during this period
  which could easily mask  the actual CME SXR front due to the flaring
  emission. For the cases where only one front is reported, the low
  cadence does seem to affect the interpretation. \citet{dai2010}
  actually observe and comment on the existence of two fronts (their
  Figure~4) but they disregard the wave interpretation on the basis of
  the low lateral speed (260\kms) compared to the speed of the CME
  (600\kms). But these are exactly the speeds expected by an MHD wave
  far from the eruption and they lack the cadence to measure the much
  higher speeds at earlier times. \citet{attrill2009} make the
  \textsl{a priori} assumption that the outer edge is the CME edge but
  they miss the outer front seen in their Figure~6(c). Therefore, we
  cannot yet find any evidence in the literature that deviates from
  the behavior described above. Namely, that the CME and the wave
  are initially cospatial and decouple once the CME lateral expansion
  begin to decelerate. We will return to this discussion in Section~11.

\begin{figure}
\centerline{\includegraphics[width=0.7\textwidth,clip=]{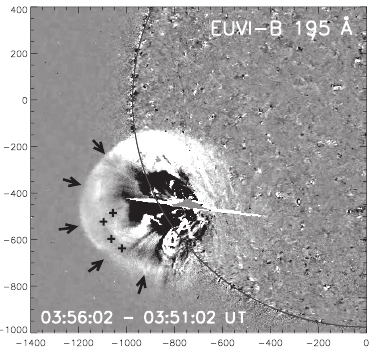}}
\caption{Observations of an EUV wave dome on 17 Janurary 2010. The
wave dome is marked by the arrows and the EUV CME by the crosses.
From \citet{veronig2010dome}. Reproduced by permission of the AAS.}
\label{fig:dome}
\end{figure}

Sometimes, the full EUV wave outline can be traced in both lateral and
radial directions.  One such example is the event on 17 January 2010
(\opencite{veronig2010dome}; \opencite{zhao2011}). The wave appeared
as a dome surrounding the erupting CME (Figure \ref{fig:dome}). The
dome was travelling faster in the radial ($\approx$ 650\kms) than in
the lateral direction ($\approx$ 280\kms) which suggested that it may
have been still driven by the CME in the radial direction whereas it
was freely-propagating in the lateral direction. The latter
interpretation was further substantiated by the constant perturbation
integral derived from the disk observations of the wave.  
  However, the difference in the observed radial and lateral speeds
  could in principle result from different fast-mode speed profiles in
  the corresponding directions.  Finally, \citet{grech2011} reached
essentially the same conclusion using a 3D blast wave model which was
in a very good agreement with the ground tracks of the wave dome both
on-disk and off-limb. Other examples of wave domes can be seen in the
events of 13 June 2010, 7 June 2011 and 4 August 2011.

Estimates on the 
the maximum lateral expansion
of CMEs in the lower corona can be derived from the recent
AIA observations of the  Kelvin-Helmholtz (KH) instability
in two eruptions
\citep[e.g.,][]{ofmankh2011, foullonkh2011}. This instability occurs at 
the interface of two fluids exhibiting velocity shear, see for example
\citet{ktkh}. In the case of the AIA coronal observations,
the two fluids are the structure of the erupting,
and rotating, flux rope and the ambient, almost potential,
coronal structures around the erupting flux rope. The instability
manifests itself in the AIA movies as the development 
of rolls at the outer boundary
of the erupting structure for an off-limb observation;
rotating core dimmings could be its on-disk manifestation.
The observations show that the KH rolls seem to be confined close
to the erupting AR which implies that
probably  the lateral expansion of the CME 
in the lower corona has a similar scale. 

The lateral extent of CMEs at the base of the corona can be also
approximated by comparing estimates of the mass content in the core
dimmings $m_{dim}$ from EUV observations and that of the associated
CME, $m_{CME}$ from WL coronagraph observations.
\citet{madimmingmass} found that $m_{dim}/m_{CME} = 1.1 \pm 0.2 $ for
a set of events observed by STEREO. Therefore, core dimmings can
supply sufficient mass to match the WL CME masses.  A
corollary of these results is that only the core dimmings could be
associated with mass evacuation. We could expect higher CME
masses than observed if the low corona opened at scales exceeding
the scale of core dimmings, as the pseudo-wave models suggest.

 \begin{figure}    
                               
   \centerline{\hspace*{0.015\textwidth}
               \includegraphics[width=0.49\textwidth,clip=]{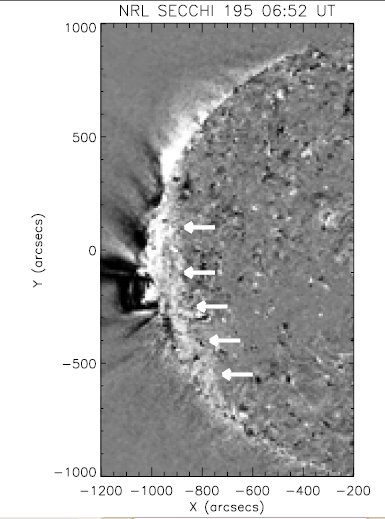}
               \hspace*{-0.01\textwidth}
               \includegraphics[width=0.49\textwidth,clip=]{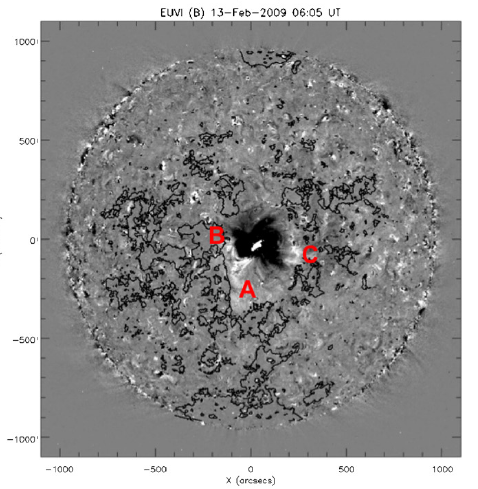}
              }
\caption{Examples of EUV waves exhibiting brightenings. Left panel from  \citet{attrill2007a}
(25 January 2007 event)
and right panel from \citet{cohen2009} (13 February 2009 event). Right panel 
reproduced by permission of the AAS.}
\label{fig:pseudo}
\end{figure}

\section{Brightenings, Secondary Dimmings and Volume Expansions}
Observations of EUV waves revealed a series of features which strongly
appeal to pseudo-wave interpretations
\citep{attrill2007a,attrill2007b,attrill2009,cohen2009,zhukov2009,dai2010,schriwave2011,warm2011}.
These include stationary brightenings, large-scale secondary dimmings
and erratic (including a series of accelerations and decelerations) or
"slow" (i.e. below the coronal sound speed) kinematic profiles like
these reported by \citet{zhukov2009} and \citet{warm2011} respectively
(see the discussion in Section \ref{sec:kin}). Such a kinematic
behavior cannot be reconciled with fast-mode waves which travel at
either constant speed or decelerate and the speed is always $\ge$ of
the fast-mode speed.

Examples of brightenings associated with EUV waves are given in Figure
\ref{fig:pseudo} where several brightenings can be seen at locations
on the EUV wave fronts for waves on 25 January 2007 and 13 February
2009. Such brightenings are seen at even higher temperatures with XRT
\citep{xrt} on
\textit{Hinode} \citep{attrill2009}.  The brightenings could result from
magnetic reconnections between the erupting flux rope and ambient QS
fields of favorable polarity. Note here that localized
  brightenings in association with EUV waves are not only seen in the
  EUV and SXRs but also in H$\alpha$ (Warmuth et al. 2004) and in He I
  10830 \AA\ \citep{vrsnak2002}.  These brightenings are seen
  slightly ahead of the EUV wave front.

\begin{figure}
\centerline{\includegraphics[width=0.8\textwidth,clip=]{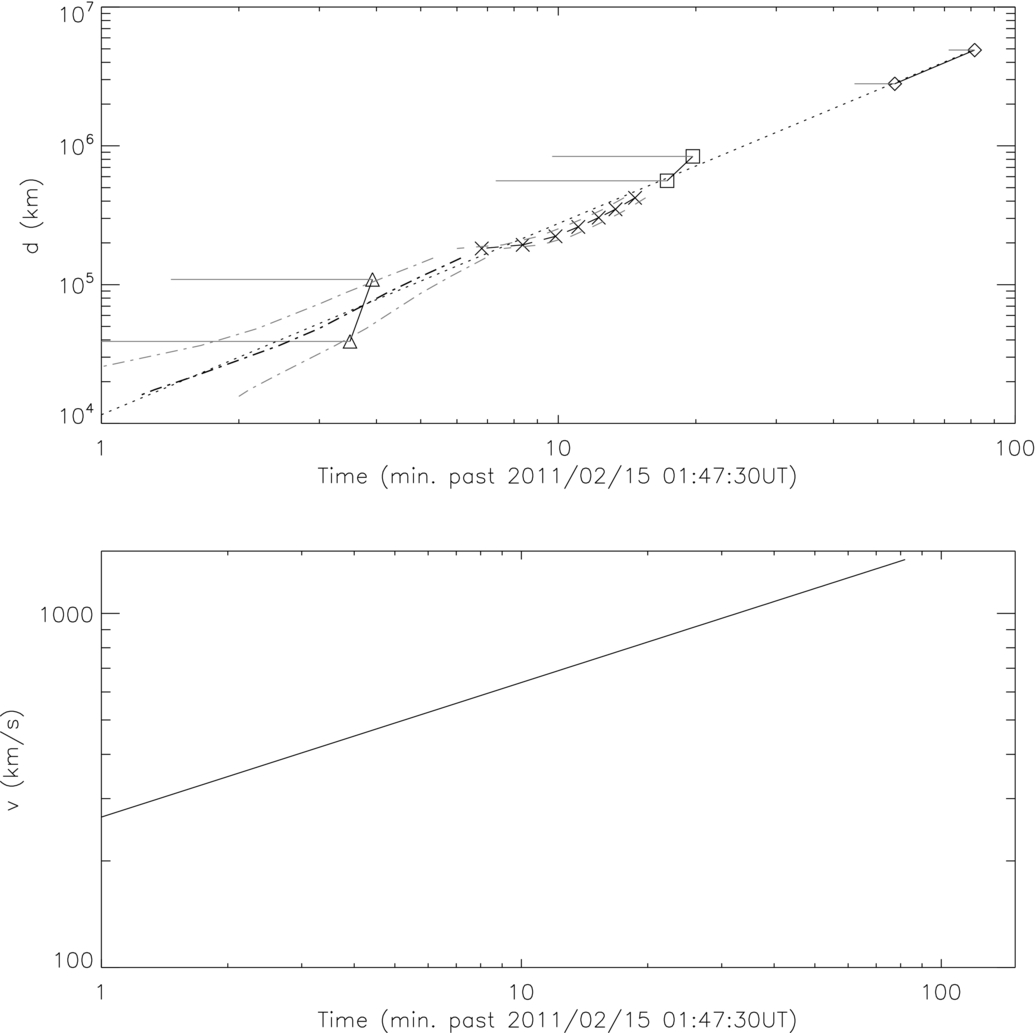}}
\caption{Observations of an EUV wave which take place on 15 February
  2011.  The upper panel contains the time-distance and the lower panel
  the speed-distance plot for several features observed in association
  with event.  For radial displacements: STEREO-A/COR2
    (diamonds), STEREO-A/COR1 (squares), STEREO-A/EUVI/195
    (triangles). For on-disk displacements: expansion front for AIA
    193Å at the central meridian measured along a great circle through
    the central flare site (crosses), expansion feature traced in AIA
    335, assuming a $45\,^{\circ}$ inclination relative to the local
    horizontal direction of the tracked loops (dash-dotted line).
  From \citet{schriwave2011}. Reproduced by permission of the AAS.}
\label{fig:volume}
\end{figure}

Smooth volume expansions are another important indicator for
pseudo-waves.  The \citet{schriwave2011} study tracked several features of the event:
the fronts of expanding loops, a bright diffuse front, and the
associated CME.  They found a smooth transition between these features,
and particularly between the expanding loops and the wave front (see
the upper panel of Figure \ref{fig:volume}). The combined kinematics
are consistent with acceleration (lower panel of Figure
\ref{fig:volume}) which under certain conditions
could be at odds with a fast-mode wave.
However, fast mode waves  can sometimes also  manifest acceleration
(see the discussion in Section 2).
Current-density renderings form an MHD simulation of the
expanding current shell around an erupting flux rope resulted in
fronts similar in appearance to the observed wave fronts.

Besides the strong core dimmings, presumably mapping to the
legs of the erupting flux rope, large-scale secondary dimmings
trailing the EUV wave front can be seen by either conveniently scaling
the images \citep[e.g.,][]{delaaul1999} or in the perturbation
profiles \citep[e.g.,][]{muhr2011}.  These dimmings could result from
the plasma evacuation behind the erupting flux rope, similarly to the
stronger core dimmings, therefore justifying a pseudo-wave
interpretation. However, plasma rarefraction in regions from where a
compressive wave has passed could have a similar effect \citep[e.g.,
][]{muhr2011}.

As discussed in the previous Sections, several of these events with
evidence of non-wave components/interpretation have been analyzed by
other groups, who found support for a wave interpretation based for
example on wave kinematics, reflections, 3D structure etc. This
however does not necessarily mean that these analyzes are are mutually
exclusive.  We will return to this very important issue in Section
\ref{sec:concl}.

 \begin{figure}    
       \centerline{\hspace*{0.015\textwidth}
               \includegraphics[width=0.49\textwidth,clip=]{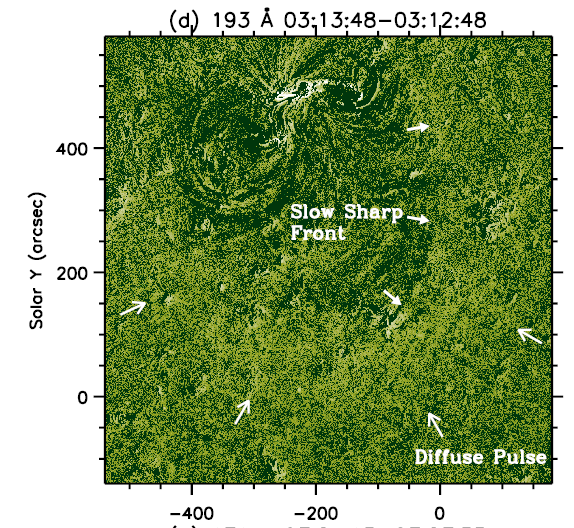}
               \hspace*{-0.01\textwidth}
               \includegraphics[width=0.35\textwidth,clip=]{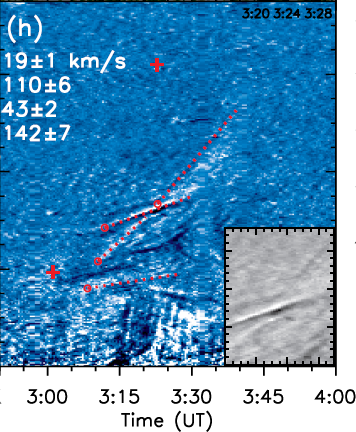}
              }
\caption{AIA observations of an EUV wave which took place on 3 April 2010.
Left panel shows the double wave front and
right panel shows multiple and crossing components in a time-distance
plot at a given direction. From \citet{liu2010}. Reproduced by permission of the AAS.}
\label{fig:liu}
\end{figure}

\section{Multiple components  and ripples}

The ultra-high cadence of AIA observations brought new information on
the structure of EUV waves.  The first observations of an EUV wave
observed by AIA were reported by \citet{liu2010}. The wave took place
on 3 April 2010 and was characterized by multiple components moving
ahead of a set of erupting loops which presumably resulted into a CME.
Besides the usual diffuse front associated with the global EUV wave,
several sharper fronts were seen moving in its wake. While the
diffuse front was moving at more or less a constant speed of $\approx$
200\kms, the two sharp fronts, a slow one at $\approx$ 80\kms\ and a faster
one at $\approx$ 160\kms, were accelerating and even crossed each other
(as seen in projection) and then propagated independently (see
Figure \ref{fig:liu}). The characteristics of the diffuse front (speed
of the order of the fast-mode speed in the QS and almost isotropic
propagation) seem consistent with a wave interpretation.  On the other
hand the sharp fronts can in principle result from compression ahead
of the expanding loops. However, this interpretation has difficulty 
explaining why after the two fronts cross they generate another set of
weaker fronts in the form of ripples. A possibility for these ripples
is that they could be related to some sort of secondary waves as found
by \citet{li2012} (see Section \ref{sec:inter}), or that they are
simply the result of oscillations of loops highly inclined towards the
solar surface.  Note here that TRACE \citep{trace} 
made the first observations of
multiple fronts associated with EUV waves
\citep{willsthomp1999,harraster2003}.  However, given the small field
of view of TRACE it was not possible to tell how far these
disturbances propagated.

\begin{figure}
\centerline{\includegraphics[scale=0.6,clip=]{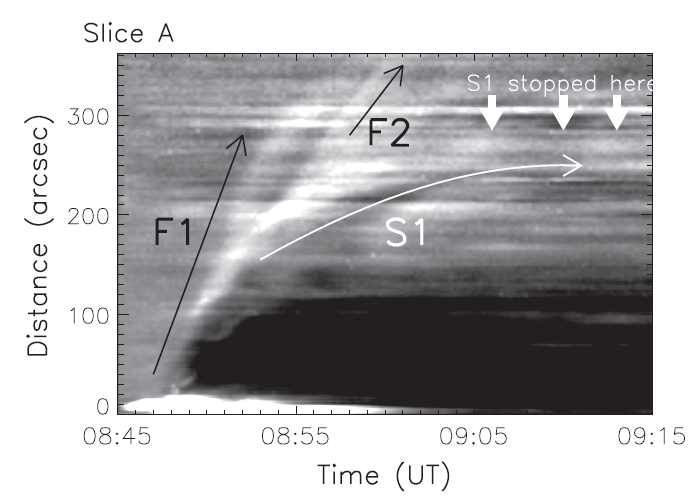}}
\caption{Distance-time plots of an EUV wave which took place on 27 July 2010 showing 
evidence of two components. From \citet{chenaia2011}.
Reproduced by permission of the AAS.}
\label{fig:chen2011}
\end{figure}

Another example of an EUV wave exhibiting multiple fronts can be found in
\citet{chenaia2011}. This event on 27 July 2010 showed evidence for two fronts. A
fast front traveling at 470--560\kms\ speeds,
followed by a slower front traveling at 170--190\kms\ speeds
(see Figure \ref{fig:chen2011}).
The slower component decelerated and eventually seemed to "stop"
at some distance from its origin. The stopping location seemed to coincide
with the location of a magnetic field seperatrix as deduced from
a magnetic field extrapolation. Another example of multiple-fronts
wave event is the 13 June 2010 event described in detail in Section \ref{sec:gener} (see
for example panel (c) of Figure \ref{fig:wavegen}).

These examples call for a hybrid interpretation
of the observations. The outer front, is consistent
with either a linear fast-mode wave  
at almost constant speed \citep{liu2010} or with a shocked fast-mode wave traveling
at higher speeds \citep{chenaia2011}. The second inner front(s) could
correspond to pseudo-waves associated with the expanding
loops of the eruption. The observed disk behavior seems consistent with off-limb observations of EUV waves
showing two fronts which split at some point (e.g., Figure \ref{fig:quad} and Section 11).

\section{Spectroscopic Observations}

Observations from the \textit{Extreme Imaging Spectrometer} (EIS; \cite{eis})
on-board
\textit{Hinode} supplied new important constraints about EUV waves.  EIS raster
observations taken over dimming regions in the cores of the ARs which
gave rise to EUV waves, showed significant blue-shifts and an increase
in the non-thermal velocity \citep[e.g.,][]{harra2007,
  imada2007,asai2008, jin2009,mc2009dimming,
  attrill2010dim,cheneis2010, cheneis2011,dolla2011}.  The observed
blue-shifts could be the signature of the radially and laterally
expanding CME.  At times, the magnitudes of the observed blue-shifts
($<$ 50\kms) are smaller than the ``typical'' propagation speeds of
EUV waves across the solar surface ($>$ 200\kms) deduced from imaging
instruments, which could be explained by the continuous inwards
bending of the magnetic field caused by the erupting flux which
decreases the line of sight component of the speed
\citep{cheneis2010}.

\begin{figure}
\centerline{\includegraphics[scale=0.6,clip=]{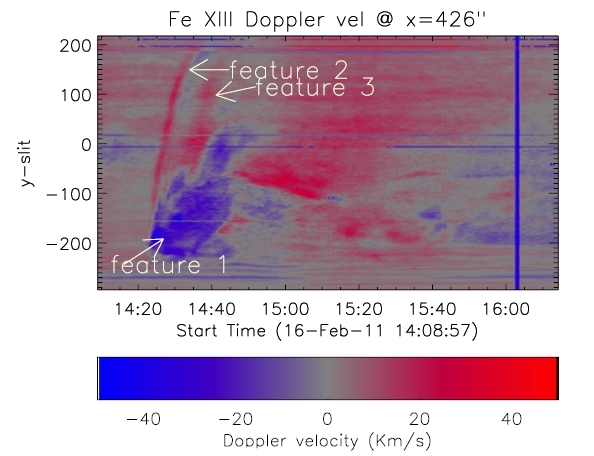}}
\caption{Spectroscopic observations of an EUV wave on 16 February 2011.
Time-distance plot of the Doppler shifts in a Fe XIII line observed
by EIS/Hinode. From \citet{harra2011}. Reproduced by permission of the AAS.}
\label{fig:harra2011}
\end{figure}

We stress here that the above observations were taken at or near the
source AR, and thus did not allow the study of the wave propagation
sufficiently far from its source. The
first observation of this kind was performed by \citet{harra2011} for an event
which took place on 16 February 2011. EIS
was taking 'sit-and-stare' observations using a 512 arcsec long slit.
The time-evolution of the Doppler-shift along the slit can be seen in
Figure \ref{fig:harra2011}.  Near the source AR (lower part of the
slit) the ``standard'' blue-shifted pattern can be seen. A
couple of outward propagating red-shifted ridges can be seen away from
the source AR.  The average speed of the ridges is $\approx$ 500\kms
which is similar to the speed of the associated EUV wave observed by
AIA.  The red-shifts could be signatures of plasma pushed downwards
and compressed by a coronal MHD shock, similar to the
\citet{uchida1968} picture for chromospheric Moreton waves. In a
follow-up study of the same event, density
sensitive line ratios of two Fe XIII lines revealed densities changes
during the wave transit along the EIS slit which were however within the noise
level \cite{veronig2011}. This supplies a spectroscopic demonstration of the small
density changes normally associated with EUV waves.  Moreover,
\citet{veronig2011} found negligible mass motions in a He II line,
suggesting that the wave was not strong enough to perturb the
underlying chromosphere.  This observation is broadly consistent with
the double-front observations discussed in the previous Section.

Spectroscopic observations could be also used
to get coronal seismology information of the
background corona where EUV waves
are propagating. \citet{west2011})
used EIS density sensitive lines to get the coronal
density over the QS where the 13 February 2009 wave propagated.
From the  propagation speed of the wave, the inferred density,
and under the assumption that 
the observed disturbance was a fast-mode wave
they obtained 0.7 $\pm$ 0.7 G for the magnetic field.

\section{Energetics}
\label{sec:ener}
Given the observationally-deduced physical parameters
of EUV  waves it is worthwhile to make some estimates
of their energy content. This is particularly important
for assessing their overall role in the energy budget of
energetic phenomena like flares and CMEs
which are associated with and occur
in tandem with EUV waves.

The kinetic energy flux $F_{kin}$ can be written as:
\begin{equation}
F_{kin}=\rho{(\delta\upsilon)}^{2}{\upsilon}_{gr}/2,
\end{equation}
with $\rho$ the mass density, ${\upsilon}_{gr}$ the group
speed and $\delta\upsilon$ the velocity perturbation.
For weak (linear) perturbations  
a lower limit for $F_{kin}$ is \citep[e.g.,][]{liu2000},
\begin{equation}
F_{kin}=\rho{(\delta\,I/I)}^{2}{{\upsilon}_{gr}}^{3}/8,
\label{eq:kin}
\end{equation}
with $ \delta\,I/I$ the relative intensity change.
It is assumed that the temperature does not change
and therefore it is $I \propto n^{2}$.
Taking  typical values for ${\upsilon}_{gr}$=300\kms,
$\delta\,I/I$=1.15 and for a QS coronal density
of $5\times{10}^{8} \mathrm{{cm}^{-3}}$
we find $F_{kin}=1.9\times{10}^{3}$\erg.

The radiative losses flux $F_{rad}$ is given by: 
\begin{equation}
F_{rad}=n^{2} L \Lambda (T),
\label{eq:rad}
\end{equation}
with $n$ the electron density,
$\Lambda (T)$ the temperature dependent
radiative losses function,
and $L$ is a characteristic lenghtscale
where the bulk of the observed 
emission comes from. We assume that $L$ does not substantially change
during the propagation of EUV waves; moreover the temperature is kept constant.
The latter assumption leads probably to an overestimate of $F_{rad}$
given that $\Lambda (T)$
is generally a decreasing function of $T$ in coronal temperatures
and sometimes there is a small temperature increase associated
with EUV waves as discussed in Section \ref{sec:thermal}.
For a 10 $\%$ increase in the density $F_{rad}$ increases
by  21 $\%$ with respect to the pre-wave conditions. Using 
the standard QS coronal radiative losses flux  from \citet{with1977}
we get $F_{rad}=1.2\times{10}^{5}$\erg. 

For coronal thermal conduction flux $F_{cond}$  it is:
\begin{equation}
F_{cond} \propto T^{7/2}/L^{2}.
\label{eq:cond}
\end{equation}
For a temperature increase of 7 $\%$
\citep{schriwave2011}
we have a 26 $\%$ increase in $F_{cond}$.
Using the standard QS coronal thermal conduction flux from \citet{with1977}
we get $F_{cond}=2.5\times{10}^{5}$\erg.

The kinetic energy flux is rather small compared
to the various QS energy terms, it can
become sizeable only for ultra-high speeds (Equation \ref{eq:kin},
for example the 2000\kms disturbance
described in  \citet{liu2000}). On the other hand,
the radiative
and conductive fluxes represent small, yet sizeable increases
over the ambient QS  corresponding values.

We can now calculate a proxy for the total enery $E_{wave}$ associated with EUV
waves:
\begin{equation}
E_{wave}=(F_{kin}+\Delta\,F_{rad}+\Delta\,F_{cond})2\pi\,R\,dR\, \Delta t,
\label{eq:energy}
\end{equation}
with the $\Delta$ quantities representing the change in the
corresponding fluxes associated with the EUV wave (i.e., the fluxes
calculated from Equations \ref{eq:rad} and \ref{eq:cond}) with respect
to the standard energy losses in terms of radiation and thermal
conduction of the QS corona \citep{with1977}. We assume that the
wave is a spherical shell of radius $R$ and thickness $dR$ and $\Delta
t$ is its life-time.  Using standard values for $R=$ 300 Mm, $dR=$ 50
Mm and $\Delta t= 40$ min we finally find that
$E_{wave}=1.8\times{10}^{29}$erg. The resulting energy is relatively
substantial; it is the energy of a small flare and lies in the low-end
of the CME energy distribution \citep{cmestat}.  This may not be
unexpected: even though EUV waves do not significantly perturb the
ambient corona they may however correspond to significant amounts of energy
given they are global phenomena.  We caution here the reader that our
energy calculations have to be seen as crude order of magnitude
estimates. 

\begin{figure}
\centerline{\includegraphics[width=0.65\textwidth,clip=]{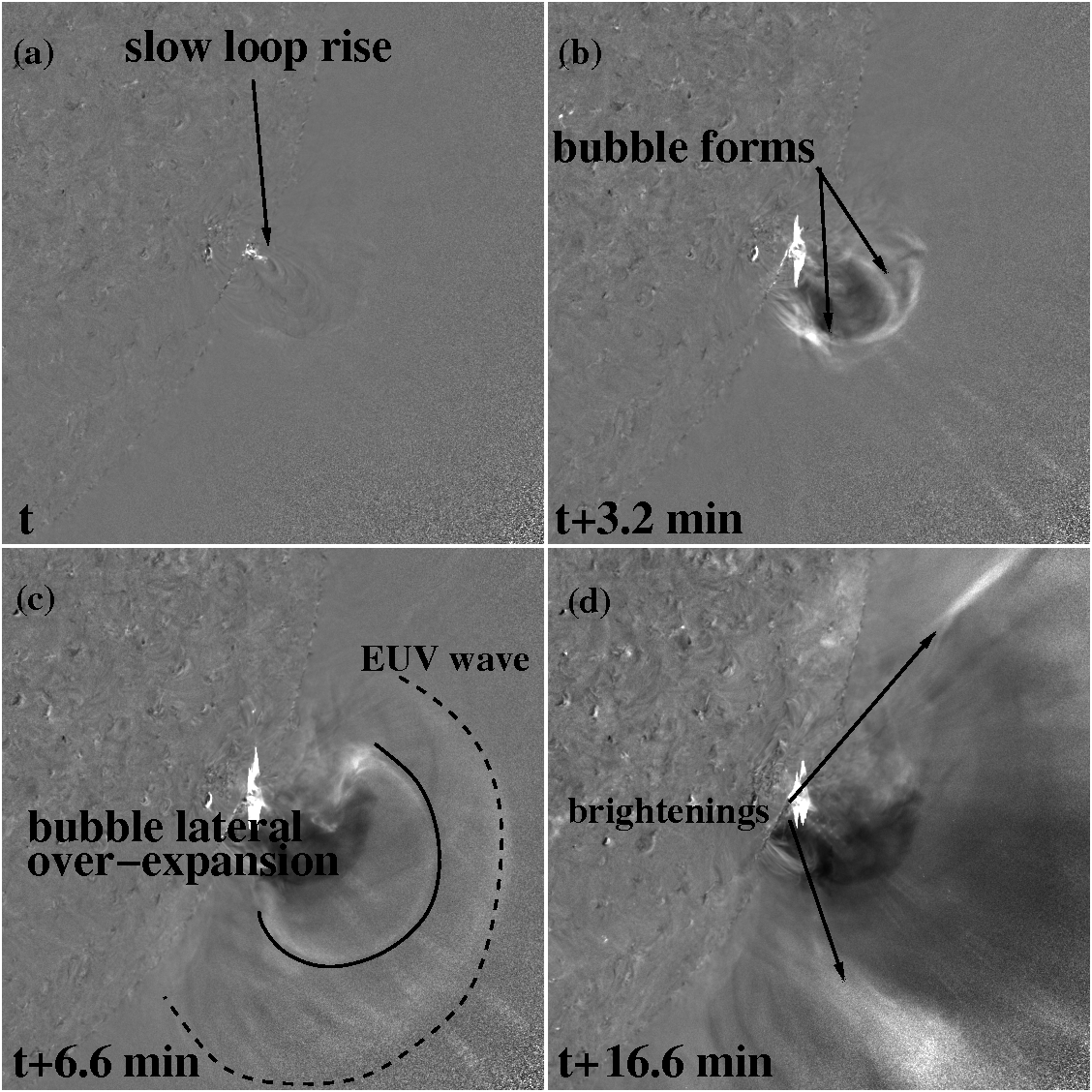}}
\caption{Evolutionary pattern towards the formation and during the EUV
wave of the 13 June 2010. Base-ratio images are shown. Slow loop rise from
the active region core (panel a) leads to the formation of a bubble (i.e. EUV
cavity; panel b). The bubble experiences a short period of lateral over-expansion
which sets a wave around it (panel c). Two quasi-stationary brightenings
form at locations where the bubble reaches maximum lateral extend (panel d).}
\label{fig:wavegen}
\end{figure}

\section{The genesis of EUV waves}
\label{sec:gener}
As discussed in the Introduction, EUV waves are associated more
with CMEs than flares. This is true from the observational as
well as from the theoretical point of view.  Indeed, most theoretical
ideas, whether wave or pseudo-wave driven, rely upon the CME as either the
wave driver or as the wave itself. High cadence movies of EUV
waves observed either on disk or off-limb show that the first
instances of a wave front are seen ahead of the flanks of the rising
and expanding loops of the early CME  \citep[e.g.,][]{long2008,
wavestereo, kienquad2009, quadwave,
bubblestereo, bubbleaia, veronig2010dome,
liu2010, ma2011, kozarev2011,schriwave2011}. Disk observations of Moreton waves show a
similar trend with the first wave fronts seen in the AR periphery and
never in its core \citep[e.g.,][]{balamoreton, temmerwavemod, muhrmoreton}.

Based on off-limb STEREO and AIA observations of impulsive events
\citep[e.g.,][]{bubblestereo, bubbleaia,
  kozarev2011,ma2011,cheng2012}, the following evolutionary pattern is
seen (Figure~\ref{fig:wavegen} and \ref{fig:wavecartoon}). A set of
loops is slowly rising in the core of the source AR region (Figure
\ref{fig:wavegen}(a)); these loops progressively start to map on the
edges of a bubble (Figure \ref{fig:wavegen}(b)), which eventually
evolves into (at least partially) the WL CME observed later with
coronagraphs. The bubble undergoes a period of strong lateral
expansion which launces the EUV wave at the flanks of the bubble;
sometimes the full wave dome becomes visible ( Figure
\ref{fig:wavegen}(c)). While the bubble reaches a more or less
constant lateral extent, the detached wave propagates further away.
The above pattern is common among impulsive events associated with EUV
waves. We have been able to gather the following partial list:
3 June 2007, 2 January 2008, 25 March 2008, 13 February 2009,
16 December 2009, 17 January 2010, 3 November 2010, 11 February
2011, 15 February 2011, 24 February 2011, and 8 March 2011.
Note that entire stages in this evolutionary pattern
would have been poorly resolved or even entirely missed without the
high cadence of the EUVI and AIA insturments.

The  expansion  of the bubble  in the radial and lateral
directions can be quantified by measuring its aspect
ratio, defined as height(t)/radius(t) of the best-fit circle or sphere to
the bubble. The aspect ratio is decreasing
for a short-period which signifies that the bubble
undergoes a period of \textit{inflation} or {\it lateral  over-expansion}, i.e.
it grows faster in the lateral than  in the 
radial direction, as was first discussed in  \citet{bubblestereo} and \citet{bubbleaia}.
This period of lateral inflation marks the launch of an EUV
wave.
The lateral over-expansion could
be driven by the high magnetic pressure within the 
bubble as it tries to reach equilibrium with the low pressure
of the ambient QS fields. 
Ideal (expansion of flux surfaces around rising flux ropes with
decreasing flux rope current) and non-ideal (reconnection adding new
flux to the erupting flux rope) MHD effects can also account for the
lateral over-expansion (Kliem et al. 2012, in preparation).  The start of
this inflationary period roughly marks the launch of the wave in their
simulations. The simulation results provide further support to the
idea that the expanding CME flanks are the trigger of the wave.

As a quantitative test for this possibility we applied the
piston-driven model of large-scale coronal waves of
\citet{temmerwavemod}. This model essentially predicts the ground
tracks of large-scale coronal disturbances given the kinematics of the
driver (the temporal evolution of its radius $r(t)$ and height $h(t)$,
for example) as well as the value of the (uniform) Alfv\'{e}n speed
($V_{A}$). Moreover, the model specifies the amplitude of the
disturbance $f(d)$ as an exponential function for example, i.e.,
$f(d)=e^{-d/p}$, with $d$ the distance from the initial position of
the driver and $p$ a scale-distance of the disturbance.

For the 13 June 2010 event, the bubble fitting
supplied the kinematics of the possible wave
driver ($h(t)$ and $r(t)$ \cite{bubbleaia}. From a polar off-limb
map of the event a $p$ of $\approx$ 0.12\rs was
deduced from the 
intensity distribution of the wave  at a height of 0.17\rs,which
is similar to the emission (formation) heights of EUV waves discussed
in Section \ref{sec:3d}. Several points along the wave 
were manually extracted from that map. Therefore, the observations
almost fully constrained the  \citet{temmerwavemod} model
with the exception of $V_{A}$. Note that \citet{temmerwavemod}
were not able
to reproduce the ground tracks of a Moreton wave which took place
on 17 January 2005 using the model above with the associated CME time-height
measurements. They proposed that either the CME flanks
or the flare blast wave generated the Moreton wave.

\begin{figure} 
\centerline{\includegraphics[width=0.9\textwidth,clip=]{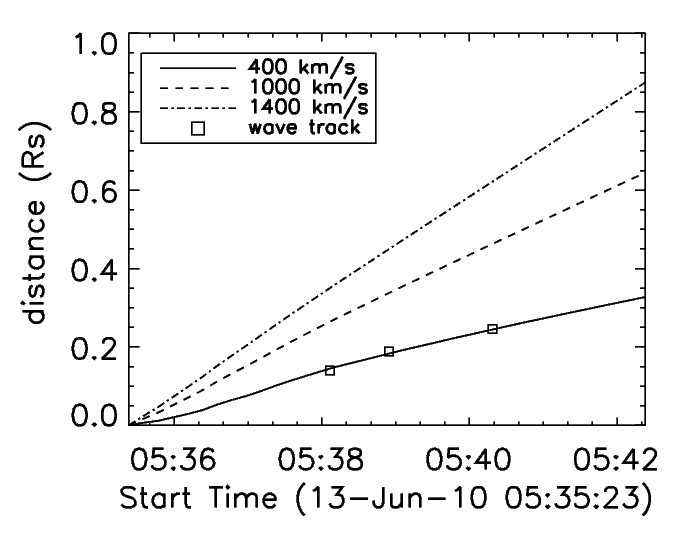}}
\caption{Predicted ground track
of a  coronal wave using the  \citet{temmerwavemod}
model. Inputs were the EUV bubble kinematics (time evolution
of height and radius) of the EUV wave which took place on 13 June 2010. 
The tracks correspond to an  Alfv\'{e}n speed of
400\kms (solid line), 1000\kms (long dashes) and 1400\kms (dash-dot);
the boxes correspond to the observed off-limb ground track of the  wave at a height
of  0.17 $\mathrm{R}_{\odot}$.}
\label{fig:ground}
\end{figure}

Figure \ref{fig:ground} contains the predicted ground tracks of the
large-scale disturbance using the observed $h(t), r(t)$, and $d$ for
different values of $V_{A}$. A very good agreement between the
predicted ground track and few selected points along the wave is
achieved for $V_{A}$=400\kms.  This value for $V_{A}$ in the low
solar corona is more appropriate to QS rather to ARs where it can
reach speeds exceeding 1000\kms. It is thus plausible that the wave
trigger were the flanks of the bubble which expanded over QS regions
rather than its front which expanded in a higher Alfvenic speed
environment.  MHD simulations show evidence of propagating
disturbances at the flanks of erupting
flux ropes (e.g., \opencite{chen2002}; \opencite{pomoell2008}; Kliem
at al., 2012, in preparation).  Further observational studies of over-expanding
bubbles are required.

\section{Towards a Coherent Picture of EUV waves}
\label{sec:concl}
EUV waves represent an excellent example of scientific endeavour. A serendipitous discovery
opens up a new area of solar physics research, leads to 
controversy over its interpretation and eventually reaches closure. Thanks
to a string of new missions and instruments over the last decade, we
are getting closer to the last stage---understanding EUV waves and their relation to
the explosive energy release in the Sun. The observations from multiple
viewpoints have played (and will continue to play) a particularly
important role in clarifying the nature of EUV waves. We believe that,
by synthesizing the existing results, we can offer a unified picture of
EUV wave signatures which effectively removes most of the wave versus
pseudo-wave controversy. This picture builds upon previous
work which considered the {\it hybrid} nature of EUV waves (see the discussion
in Introduction).
Before we discuss this picture, it is
useful to briefly recap the most important findings from the
recent observations. In the following
list, we mark each item with a \w and/or \pw to show whether it is
consistent with a wave or a pseudo-wave interpretation, respectively.
Table~1 contains a summary of some properties of EUV waves.
\begin{itemize}
\item Despite a large range of initial speeds, all EUV waves
  decelerate to a narrow range of $\sim$200-300\kms\ which corresponds
  to the nominal fast-mode speed in the quiet Sun (\w).
\item Observations of wave reflection and transmission at coronal hole
  boundaries (\w).
\item Observations of decelerations or complete disappearances in ARs
  (\w).
\item EUV wave and CME kinematics differ (\w).
\item Temporary dimmings (\w).
\item Long-term dimmings (\pw). 
\item Brightenings at the wave front and its wake (\pw).
\item Multiple wave fronts travelling at different speeds. Sometimes
  they cross each other and produce secondary ripples (\w and \pw).
\end{itemize}

The observations (and MHD modeling) suggest that global EUV waves are
more consistent with a (fast-mode) wave interpretation. This is rather
expected since any kinetic disturbance will launch waves in a
magnetized plasma. It is not surprising, therefore, that the fast-mode
wave was the initial interpretation put forth soon after the discovery
of this phenomenon \citep{thomp1998, thomp1999}.  Then, why does the
controversy over the nature of the EUV waves persist? Why are there
cases where different observers reach opposite conclusions from the
analysis of the very same events?

We believe that the answers to these questions lie in a sort of 'structure
confusion'. EUV waves, being associated with CMEs, occur in tandem with an
extensive 'zoo' of other phenomena (e.g., core and
extended dimmings, stationary brightenings, deflections and
oscillations, multiple fronts and ripples, flows, etc). All these
phenomena evolve in time and size and hence can be associated, or rather
confused, with the wave. The confusion can be avoided if the definition of the EUV wave is kept in mind. In other words, the EUV wave is \textit{the outermost propagating intensity front reaching
  global scales}. Most of these phenomena extend, at best, to nearby
ARs or coronal holes. It is therefore important to carefully trace the
proper front (the outermost one) amidst the multitude of all the other
structures and time-evolving phenomena. In that case, combinations of disk and off-limb observations 
as well as full-sphere viewing and high cadence can prove invaluable.

\begin{figure}
\centerline{\includegraphics[scale=0.8,clip=]{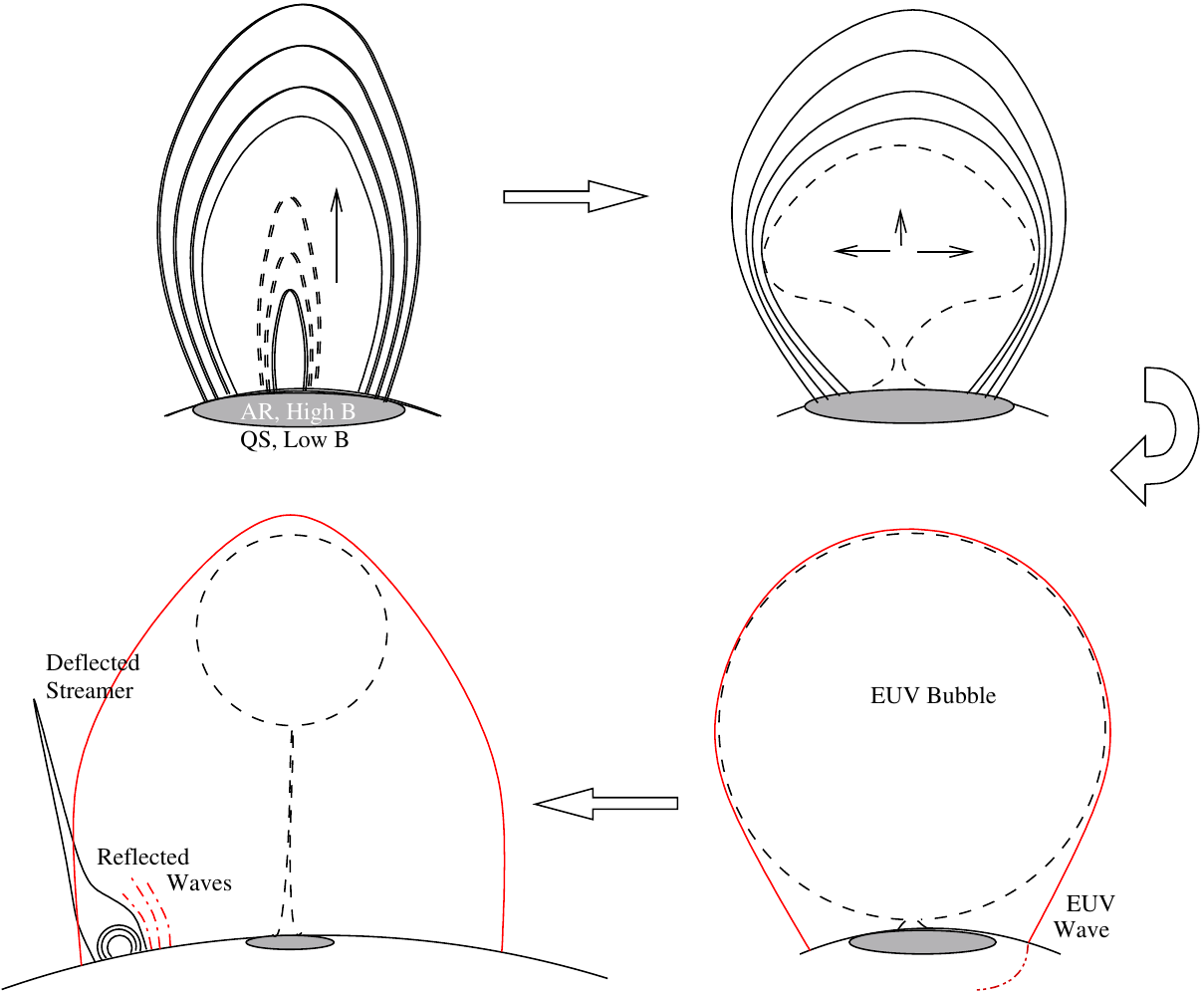}}
\caption{Schematic evolution of the genesis of a CME and its
  associated EUV wave.  The cartoon represents a top-level synthesis
  of EUV wave observations as discussed and interpreted in this
  review.  Top left: A set of outward expanding loops in the core of
  an active region (grey area) always occur before a CME-EUV wave
  eruption. Top right: At some point, the loops disappear and give
  rise to an expanding bubble.  The bubble undergoes a very short
  period of \textit{fast lateral expansion\/} compressing the
  surrounding field as it expands. Bottom right: The bubble eventually
  reaches its maximum lateral extent in the low corona and grows
  self-similarly from that point on. The fast expansion drives a wave
  around the edge of the EUV bubble.  The wave forms when the lateral
  expansion enters the low magnetic field of the Quiet Sun (where the
  fast mode speed is lower than in the AR). The wave is driven at this
  stage with potentially a very small standoff distance.  The wave
  \textit{decouples} from the CME when the lateral expansion speed
  drops. Bottom left: At a later stage, the CME is moving out in the
  corona, possibly driving a shock around it. The original EUV wave is
  propagating at larger distances reflecting off and/or propagating
  through ARs or coronal holes, and causing deflections and
  oscillations in EUV loops. It may be no longer driven. 
See Figure~\ref{fig:wavegen} for an actual observation that corresponds closely to this scenario.}
\label{fig:wavecartoon}
\end{figure}

The next step, is to realize that the terms 'EUV wave' and 'CME' are
not equivalent. In our experience, this is the main source of
confusion regarding the EUV wave nature in the literature. Sometimes the 'EUV wave' is
singled out while discussing the CME front or vice versa. Other times
attempts are made to connect white light fronts observed with
coronagraphs to fronts observed with EUV instruments but not at the
same height and finally other times there are "jumps" from wave to CME
front as an event is followed in time. The root of the problem seems
to be the imprecise definitions used sometimes for these terms. As
part of our synthesis effort, we propose a set of definitions here: an
EUV wave is a disturbance propagating along the EUV coronal surface at
large distances from an eruption. A CME is a magnetic disturbance
(i.e. a magnetic fluxrope) propagating outwards from the corona. 

The strong association of EUV waves to CMEs implies that an
\textit{expanding driver\/} is likely following the expanding EUV wave
(Figure~\ref{fig:wavecartoon}). Both structures will appear as a wave
front in the images but their behavior (and interpretation) will be
different. For example and from the discussion of Section~2, the wave
will propagate at the local fast-mode speed when it is in the
  linear regime or close to it and above it when it is in the
  non-linear regime while the CME can propagate at any speed. The
wave will propagate to $> 60^{\circ}$ from the erupting region while
the CME flanks will stop at a coronal hole boundary or within
$30^{\circ} - 40^{\circ}$ since the average width of CMEs is
$\sim60^{\circ}$ \citep{stcyr00,yashiro04}.

By taking into account the dual nature of EUV wave/CME, most of the
controversy over the nature of EUV waves can be removed. For example,
for events showing evidence of two wavefronts (e.g. Figures
\ref{fig:liu}, \ref{fig:chen2011}, \ref{fig:wavegen}), the inner
brighter front is the expanding CME loops or bubble (and hence the
\textit{pseudo-wave}). The outer fainter front is the fast-mode wave
(and hence the \textit{wave}). The wave can be driven or
freely-propagating depending on the stage of the CME evolution. These
events are characterized by the existence of an EUV bubble or of a
well-formed white light CME fluxrope \citep{quadwave, bubblestereo,
  liu2010, veronig2010dome,
  kozarev2011}. 

Other events lack a
well-defined CME structure in the EUV corona. But the CME is there and it will
cause 'pseudo-wave'-like signatures. These are the cases that lead to opposite interpretations of the same data when the event is
analyzed partially. An example could be the 15 February 2011
event. The relatively smooth transition among the height-time plots of
the EUV wave, loops and white light CME reported in
Figure~\ref{fig:volume} does not imply that the EUV wave and the CME
front are the same (and hence that the EUV wave is a pseudo-wave as
suggested by \citet{schriwave2011}) but simply that the EUV wave and
the CME front are {\it at the same location}. Without a clear manifestation
of a CME front in the EUV it is hard to discount the latter
interpretation. However, the reported volume expansions 
have, at some point, launched the "true" EUV wave undergoing reflections/transmissions
as shown in \citet{oo2012}.

Distant brightenings are commonly discussed in
such events in support of a pseudo-wave interpretation. Their
explanation in terms of the CME driver is rather straightforward. When
the erupting flux reaches its maximum lateral extension in the low
corona it can generate features pertinent to pseudo-waves like
stationary or moving brightenings. For example, strong brightenings at
the sides of the erupting flux can be easily discerned from a side
view (panel (d) of Figure \ref{fig:wavegen}). These are formed when
the erupting flux ``stops'' at those locations (coronal hole
boundaries) and ambient plasma is compressed and maybe heated. Given
the inclination of these compressed structures, a disk observation of
such an event would give the impression of moving brightenings
\footnote{STEREO A observations with a disk view of this event did not
  have enough cadence to capture these features.}. A similar situation
could have occurred, as discussed before, in the event analyzed by
\citet{chenaia2011}, see Figure \ref{fig:chen2011}, where a stationary
brightening forms where the inner wave front seems to ``stop''.

We therefore propose that the majority of the observations can be
reconciled by properly considering the spatial and temporal evolution
of the CME vis-\`{a}-vis the expanding EUV
wave. Figure~\ref{fig:wavecartoon} is our attempt to provide
  a unified picture of the CME-EUV wave coupling. It is a simplified
  cartoon which emphasizes the most important, top-level
  characteristics of the eruption but it can still account for the
  various observations of waves (or pseudo-waves). Figure~\ref{fig:wavegen} is the 
closest observational example we could 
 find.  In this picture,
the EUV wave is driven by an expanding CME. The wave appears if or
when the speed of the CME expansion overtakes the local fast-mode
speed. This happens more easily at the flanks since they encounter QS
conditions (and hence lower fast-mode speeds) sooner than the radially
outward moving CME parts. The wave front will tend to be very close to
the CME front (small standoff distance) for fast accelerating
CMEs. The two will separate when the CME either slows down or the wave
encounters a lower fast-mode speed environment. There is no guarantee
that an EUV front will form at the CME nose if (or while) the CME
propagates inside a streamer because of its high plasma $\beta$. This
implies that connecting a white light front to an EUV front must be
done very carefully. As far as we can tell, this picture explains all
of the available observations of EUV waves and contains both the wave
and pseudo-wave interpretations.

\begin{figure}                  
   \centerline{\hspace*{0.015\textwidth}
               \includegraphics[width=0.49\textwidth,clip=]{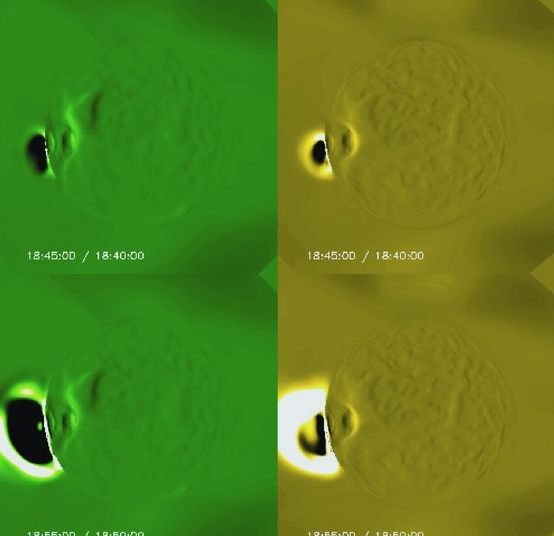}
               \hspace*{-0.01\textwidth}
               \includegraphics[width=0.49\textwidth,clip=]{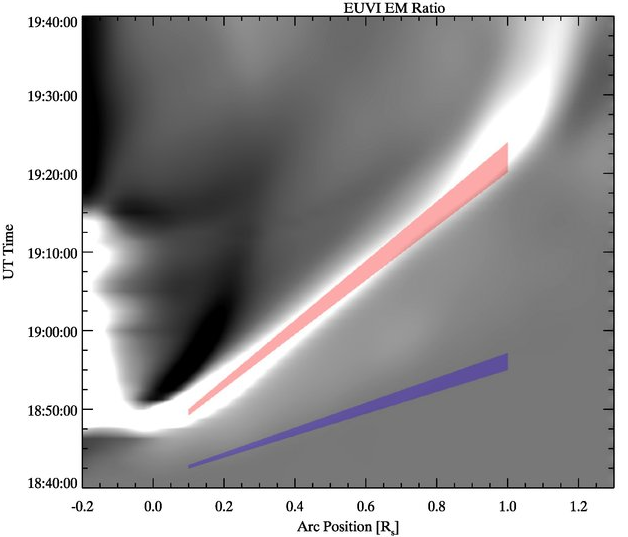}
              }
\caption{Global MHD simulations of an EUV wave which
took place on  25 March 2008. Left panel: a series of synthetic 
base-difference images for the 195 (green) and 284 (yellow) EUVI channels.
Right panel: time-distance plot of emission measure along
a sector.  From \citet{downs2011}. Reproduced by permission of the AAS.}
\label{fig:downs}
\end{figure}

Furthermore, this picture is strongly supported by MHD modeling. The recent
advances in global MHD modeling of EUV waves recover many of the
observed signatures \citep{linkerwave, lion2009,cohen2009, 
schmidt2010,downs2011}.  
These
models have the ability to generate realistic synthetic images of the
global solar corona before and during EUV waves.  This is achieved by
solving consistently the field-aligned energy transport and using
realistic boundary conditions. The waves are usually driven by the
eruption of a postulated flux rope or by simply launching a velocity
pulse.  Several features seen in the observations like bubbles,
multiple fronts, reflections \textit{etc} can be seen in these simulations. As
an example Figure \ref{fig:downs} shows results from a recent study of
the 25 March 2008 event \citep{downs2011}.  The simulated EUV images
of the left panel of this figure reproduce the bubble-like structure
and propagating intensity fronts. They have similarities with actual
observations shown in, for example, Figures \ref{fig:over} and
\ref{fig:dome}.  The right panel of Figure \ref{fig:downs} shows
time-distance plots of simulated emission measure along a track along
the wave path.  Two lanes can be seen, which resembles plots with
similar format for observed events (e.g., Figure \ref{fig:chen2011}).

We note here that our picture may not be able to explain every nuanced
structure associated with EUV waves (e.g., see the crossing ripples
in Figure \ref{fig:liu} and the possible
explanations discussed in Section 7). But one needs to keep in mind that EUV waves
do not have to be related to all observational features seen during
explosive events. Therefore, the latter should not be included in
the interpretations of such phenomena. This becomes evident in the
most dramatic way with the ultra-high cadence and high sensitivity AIA
movies where the entire corona seems to be in a stage of seamless
agitation at any time including when EUV waves take place. An example
of non-related features during EUV waves is narrow angular extent
intermittent flows firing from the core of the source ARs {\it after}
the early CME is underway. These are
probably  triggered by the general reconfiguration of the coronal
magnetic field during a CME but have nothing to do with the wave
itself.

%

\begin{table}
\caption{Properties of global EUV waves compiled from recent observations.}
\label{table:table1}
\begin{tabular}{ll}     
\hline
average speeds (\kms) & 200-400 \\
initial speeds (\kms) & 223-750 \\
final speeds (\kms)   & 180-380     \\ 
maximum distance from source (Mm) &  350-850  \\
width (Mm)            &    20-250  \\   
acceleration  (\acc) & 2.0$\times{10}^{3}$- 150 \\
magnetosonic-Mach number & 1.08-1.4 \\
intensity increase ($I/I_{0}$) & 1.02-1.7 \\ 
maximum temperature (MK)  & 2.8   \\
maximum density increase     & 1.009-1.3 \\
emission heights         &   50-100 Mm \\
3D shape                 & $\approx$ spherical \\
relationship with CME    & co-spatial (early), ahead (later) \\
kinetic energy flux (\erg)  &   1.9$\times{10}^{3}$      \\
radiative losses flux (\erg)  &   $1.2\times{10}^{5}$     \\         
thermal conduction flux (\erg)  & $2.5\times{10}^{5}$    \\
\hline
\end{tabular}
\end{table}

\section{Future areas of work} \label{sec:future}
We have tried to present the large amount of recently published work on EUV
waves and to suggest a simple scenario for their genesis which reconcile the sometimes
conflicting interpretations of their observations. In our view, the
community is not far from the resolution of the nature of EUV waves.  
However, important questions remain unanswered. Research will benefit greatly
from the availability of high cadence observations from AIA and the
full-Sun coverage afforded by the combination of AIA and EUVI-A and -B
since mid-2011. We discuss below a list of such open questions.

\begin{enumerate}
{\it \item What determines the generation of a true wave?}
We strongly suspect that the {\it impulsiveness} of the early CME
evolution is the key parameter.  Theoretical studies show that the
driver needs to accelerate in a matter of few minutes to launch 
a large-scale wave or a  shock with pulse amplitudes sufficient for
detection \citep[e.g.,][]{zic, pomoell2008, downs2011}.  
The peak speed seems not to be the main factor,
as long as it exceeds (at some point) the characteristic speed of the
ambient medium. We have observed EUV waves in association with
relatively slow drivers; e.g., the 13 February  2009 event with a
radial speed of around 200\kms (c.f. Figure \ref{fig:quad}) and the
13 June 2010  event with a peak speed of 400\kms and rapid
deceleration. The similarity in both events is the sharply peaked
acceleration profile. Events with similar speeds but with more gradual
acceleration profiles are not prone to generate visible EUV waves. The
lack of observable wave signatures does not necessarily mean lack of
waves. All motions in a plasma will drive some sort of a wave
disturbance. However, gradually accelerating drivers will tend to
create lower amplitude pulses farther away from the eruptions
site. Hence, the associated wave density enhancements will be weaker
and may escape detection. Deflections of distant (from the eruption
site) structures are as common in EUV images as they are in
coronagraph images. 
The corollary, for gradually accelerating events, is that any
propagating EUV signatures will be associated with the expanding CME
(the pseudo-wave) and will be likely restricted to long-term dimmings
in the vicinity of the eruption.

We have suggested in the past, and discussed in
Sections 10 and 11, that the strong lateral expansion is the key factor for the
wave generation.  Therefore, the determination of the duration and the lengthscale
of the driver acceleration in both the radial and lateral direction
for several EUV wave events should be an important task for the
future.

{\it \item Determine when and where pseudo-wave and wave decouple.}
The ambient environment is probably a significant player in this
question.  In Section \ref{sec:gener}, we discussed how the expanding
CME bubble, and more precisely its flanks, launch an EUV wave.  The
flanks, as they encounter weak QS fields, will strongly expand. The
expansion will happen closer to the source AR, and occur over a larger
area, when the AR is surrounded by mainly QS areas, which is the case
during solar minimum conditions.  However, as the cycle peaks, nearby
ARs and equatorial coronal holes may prevent the strong lateral expansion of the CME bubble or the
formation of a large scale wave. In this case, we expect that the
volume expansion of the bubble, i.e. the pseudo-wave, would dominate
the wave signatures until the true wave forms further away (assuming a
sufficiently energetic event). Therefore, EUV waves
taking place during solar maximum conditions could have more
pronounced CME (pseudo-wave) signatures.  

{\it \item Establish the exact relationship between type-II metric shocks
  and EUV waves.}  
Statistical studies show a rather high degree of correlation
between metric type-IIs and EUV waves. 
\citet{biese2002}  showed that metric type-IIs occurred in tandem
with at least 69$\%$ 
of EIT waves and 
\citet{klassen2000}
found an even higher degree of correlation (almost 90\%).
\citet{warm2004} found a 100\% correlation between
type-IIs and Moreton waves. Finally, radioheliographic observations
show that the sources of type II burst are generally consistent with coronal
wave signatures \citep[e.g.,][]{pohjo2001,khan2002,vrsnak2006}.

Even though the existence
of a close association between metric type-IIs
and EUV waves is well established some of the details are missing.
Metric type-IIs generally last from three to ten minutes
while the EIT cadence is 12 minutes or so. The scarcity of metric
type-IIs during the extended minimum prevented any good comparisons
with EUV observations but the situation has changed in the last
years. \citet{kozarev2011} and \citet{ma2011} presented detailed
  comparisons of EUV wave and metric type-II kinematics showing a
  close correspondence between the two for the 13 June  2010 event;
  namely, the start of the radio emission start and the EUV wave
  appearance coincide to within less than a minute.  
They estimate the
  origin of the radio emission at the nose of the outgoing CME/wave. 
The joint 
 analysis of the band-splitting of the radio dynamic spectrum
 and the stand-off distance between the shock driver (bubble) and
the shock from the EUV observations of this event allowed to infer
some estimates of the magnetic field at the base of the corona ($\approx 1.3-1.5$ G, \opencite{gopal13june}).
A study of several more events by \citet{vour2011} suggests that
  the radio emission appears at, or near, the peak of the CME
  acceleration profile when an EUV wave also forms. In most cases, the
  EUV wave is visible only at the flanks of the CME, so these authors
  proposed that the radio emission originates at the flanks of the
  outgoing CME, in agreement with past imaging results from metric
  type-IIs \citep{gary84}. It seems, therefore, that the radio
  observations provide strong support for the wave nature of EUV waves
  but the origin of the metric type-II emission (front or flanks)
  remains unclear. This is an important area of research for 
the future.

  {\it \item How frequent are EUV wave reflections?}  We saw in
  Section \ref{sec:inter} that sometimes EUV waves seem to reflect at
  or even to go through coronal holes and ARs. This is the expected
  behavior and should be common, if EUV waves are true waves. But to firmly establish this, we
  need to expand from individual event to statistical studies.
  Perturbation-profiles analysis can be used to test if energy is
  roughly conserved between the incoming and the reflected and
  transmitted waves.

  {\it \item 3D relationship between EUV waves and CMEs.}  A limiting
  factor in studies relating EUV waves and CMEs in 3D of Section
  \ref{sec:3d} is that they are dealing with slow associated CMEs. As a result,
  when the CME emerges into COR1 field of view the EUV wave is weak
  and diffuse. It is thus desirable to perform this type of CME-EUV
  wave modeling/comparison for EUV waves which are associated with
  fast CMEs. This will allow the CME to emerge early enough into the
  coronagraph field of view when the wave is still
  strong. PROBA-II/SWAP\citep{swap}  off-point EUV images which can follow the
  early EUV CME at much larger heights than any current EUV imager, up
  to 1 \rs above the solar limb, can also help into the wave-CME
  comparisons.

{\it \item What is the energy budget of EUV waves?}  In Section
  \ref{sec:ener} we made some order of magnitude estimates of the
  energy content of a 'typical' EUV wave. It is highly desirable to
  perform detailed calculations of the various energy terms for
  specific EUV wave events as a function of time and then deduce the
  total energy shed into each wave event. The wave energy could be compared with the
  corresponding energies of the associated CMEs and flares to determine the energy partition.  AIA
  multi-channel coronal observations can supply DEMs at each point
  across and during EUV waves which can be then used to determine the
  corresponding radiative losses.

  {\it \item Establish possible links between Solar-Energetic-Particles (SEP)
  events and EUV waves.}
Recent work  by \citet{rouillard2011} and
Rouillard et al. (2012) (submitted)  combine the 360-degree coverage of EUVI with AIA wave observations and shows that 
the lateral extensions of strong (i.e., associated with
shocks) EUV waves could provide reliable estimates on the
angular extensions, injection times, and intensities and of associated SEP events at the coronal base. Such work 
supply important clues on the origins of accelerated particles
in SEP events (i.e. low \textit{vs.} outer corona origins).

  {\it \item What is the role of flares as drivers of EUV waves?}  The
  strong increase in plasma pressure associated with flares is capable
  of launching a blast-wave.  While it seems as if flares are quite
  unlikely to drive global EUV waves (see the discussion in the
  Introduction) they may be a wave-driver under some special
  conditions. The key here may be the spatial location of the
  flare. If the flare occurs at or close to the AR core, the very
  strong vertical gradients in the fast-mode speed would strongly
  refract the wave upwards with little lateral expansion 
  \citep{wang2000,wavestereo}.  On the other hand, waves
  launched with some offset from the AR core would expand more easily
  in the lateral direction and therefore give rise to a global EUV
  wave.  Moreover, calculations of the volume expansions of flare
  heated loops showed that they can drive large-scale waves and shocks
  if the flare occurs away from regions of very low magnetic $\beta$
  (i.e. AR cores, \opencite{vrsshockrev}).  Another important parameter
  is the timing between the flare and wave-onset. For example
  \citet{muhrmoreton}, \citet{wavestereo} and  \citet{veronig2008wave}
  found that the associated flares peaks occur before the wave onsets
  thus invalidating a flare driver. Therefore, statistical
  studies of the flare locations, for particularly wave events without
  an associated eruption, and of the relative flare-wave timings are
  required.

{\it \item Establish the relationship between Moreton waves and EUV
  waves.}  Previous studies showed that if Moreton waves are the
chromospheric counterparts of coronal waves observed with EIT, then
the latter should experience significant deceleration during
their early stages
\citep{warm2001}. However, the low EIT cadence 
(12 minutes versus 1-2 minutes of the H$\alpha$ observations)
did not allow to
directly check that in full detail. Note here
that the majority of the EUV waves associated with  Moreton waves 
in \citet{warm2010}
shows evidence of deceleration (this could of course only 
be determined for events with EUV fronts visible in more than two images).
Similar behavior is seen in SXR observations of coronal waves,
when available.
In cases for which the early EUV wave deceleration could have
missed probably due to low cadence, the extrapolated Moreton
wave track roughly matches that of the EUV wave later on
during several events. 
These results suggest a strong relationship between 
Moreton waves and EUV waves.

The ultra-high AIA cadence allows for much more
detailed comparisons between Moreton and EUV waves
by combining AIA and H$\alpha$ or He I 10830 \AA \,
observations, which will supply more detailed
kinematics of both phenomena. The first example
of such a comparison is reported in \citet{asai2012}
where cospatial  H$\alpha$ and EUV fronts
were dedected.

Note that AIA observations in the flare channels (94 and
131) may also help tracing EUV waves into lower temperatures. These
channels, apart from their main peaks at very high temperatures, they
have secondary weaker peaks at transition region
temperatures. Therefore, if and whenever there is a wave extension
towards lower temperatures, this could be searched by adding together
several 94 or 131 \AA~ images to bring up the signal to noise ratio.

{\it \item Establish the properties of off-limb wave-related
    oscillations.}  These oscillatory phenomena are an excellent
  'smoking gun' for the action of EUV waves. We need to fully
  characterize and map their properties, like period, amplitude,
  damping time etc, like done for individual oscillation events 
  \citep[e.g.,][]{ma2011kink}.  Determining in particular the initial phase of
  these oscillations and whether different oscillating structures are
  sharing or not the same initial phase could help into discriminating
  between different possibilities for their EUV wave origin
  (see Figure \ref{fig:defle}). Such analysis will also allow to infer
  detailed coronal seismology information over the large areas over
  the Sun that EUV waves  are propagating. Moreover we can assess their
  contribution to the decay and the energetics of EUV waves.

\end{enumerate}

\begin{acks}
  We thank the referee for the very useful comments. S.P. acknowledges
  support from an FP7 Marie Curie Re-integration Grant
  (FP7-PEOPLE-2010-RG/268288).  He also thanks the Scientific
  Organising Committee of the "Sun-360: Stereo-4/SDO-2/SOHO-25
  Workshop" for an invitation to give a talk on EUV waves which is the
  base of the present review. A.V. is supported by NASA contract
  S-136361-Y to the Naval Research Laboratory.
\end{acks}

%

%

%

%
%

\bibliographystyle{spr-mp-sola-cnd} 
\bibliography{bib_sol.bib}

\end{article} 
\end{document}